%% file: susy_draft_8.tex
\documentclass[aps,preprint,nofootinbib,preprintnumbers,eqsecnum,superscriptaddress]{revtex4-1}

\input HLdefs

\newcommand{\local}{{\rm dynamical}}

\allowdisplaybreaks

\begin{document}

\title{%Dynamical KMS symmetry and 
Emergent Supersymmetry in Local Equilibrium Systems}

\preprint{MIT-CTP/4861}

\author{Ping Gao}
\affiliation{Center for the Fundamental Laws of Nature, 
Harvard University, 
Cambridge, MA 02138}

\author{Hong Liu}
\affiliation{Center for Theoretical Physics, \\
Massachusetts
Institute of Technology,
Cambridge, MA 02139 }

\begin{abstract}
Many physical processes we observe in nature involve variations of macroscopic quantities over spatial and temporal scales much larger than microscopic molecular collision scales and  can be considered as in local thermal equilibrium. In this paper we show that any classical statistical system in local thermal equilibrium  has an emergent supersymmetry at low energies. 
We use the framework of non-equilibrium effective field theory for quantum many-body systems defined on a closed time path contour and consider its classical limit. Unitarity of time evolution requires introducing anti-commuting degrees of freedom and BRST symmetry which survive in the classical limit.  The local equilibrium is realized through a $Z_2$ dynamical KMS symmetry.  We show that supersymmetry is equivalent to the combination of BRST and a specific consequence of the dynamical KMS symmetry, to which we refer as the special dynamical KMS condition. In particular, we prove a theorem stating that  a system satisfying the special dynamical KMS condition is always supersymmetrizable.  We discuss a number of examples explicitly,  including model A for dynamical critical phenomena, a hydrodynamic theory of nonlinear diffusion, and fluctuating hydrodynamics for relativistic charged fluids.

\noindent

%We develop an effective field theory for dissipative fluids which governs the dynamics
%of gapless modes associated to conserved quantities.  The system is put in a curved
%spacetime and coupled to external sources for charged currents. The invariance of the
%hydrodynamical action under gauge symmetries and diffeomorphisms suggests a natural
%set of dynamical variables which provide a mapping between an emergent ``fluid spacetime''
%and the physical spacetime. An essential aspect of our formulation is to identify the appropriate
%symmetries in the fluid spacetime.
%Our theory applies to nonlinear disturbances around a general density matrix.  For a thermal
%density matrix, we require an additional $Z_2$ symmetry, to which we refer as the local KMS condition.
%This leads to the standard constraints of hydrodynamics, as well as a
%nonlinear generalization of the Onsager relations. It also leads to an emergent
%supersymmetry in the classical statistical regime, with a higher derivative version required for the full quantum regime.

\end{abstract}

\today

\maketitle

\tableofcontents

\section{Introduction}

The goal of many-body physics is to explain and predict macroscopic phenomena.
Except for  some very simple systems, however, it is rarely possible to  compute macroscopic behavior  of a system directly
from its microscopic description. 
% as microscopic degrees of freedom 
%are often unsuitable/inconvenient for describing macroscopic phenomena which are collective in nature.  
For static properties of equilibrium systems, % which can be extracted from equilibrium partition functions, 
we have the extremely successful Laudau-Ginsburg-Wilson paradigm, which 
provides an effective field theory (EFT) description of long-distance (IR) physics 
\be \label{par0} 
Z [\phi]= e^{- \beta F [\phi]} = \Tr e^{-\beta H } = \int D \chi \, e^{- S_{\rm eff} [\chi; \phi]}  \ .
\ee
In~\eqref{par0}, $\phi$ denotes collectively external sources, $\chi$ denotes collectively gapless modes, and $S_{\rm eff}$ is the  low energy effective action of the gapless modes obtained by integrating out gapped degrees of freedom. %The path integrals over $\chi$ in~\eqref{par0} may be considered as resulting from  up to cutoff $\ep$.
While in practice such direct integrations are almost always impossible, one can deduce the general form of $S_{\rm eff}$ on physical ground.  Two elements are needed for this purpose: (i) choice of the IR dynamical variables $\chi$ which best capture gapless (collective) degrees of freedom; (ii) symmetries of $S_{\rm eff}$.  
One can then write down $S_{\rm eff}$ as the most general {\it local} field theory  consistent with the symmetries.
 %Two elements are needed for this purpose: 

For a non-equilibrium system or dynamical quantities of an equilibrium system, partition function is inadequate. A large class of non-equilibrium observables can be extracted from the generating functional defined on a 
closed time path (CTP)~\cite{schwinger,keldysh,Feynman:1963fq} %for reviews, see~\cite{Chou:1984es,Hubook,Kamenev}), 
\bln \label{gen0}
e^{W [\phi_1, \phi_2]}  & = \Tr \le( U (+\infty, -\infty; \phi_1) \rho_0 U^\da (+\infty, -\infty; \phi_2) \ri) \\
& =   \int_{\rho_0} D \psi_1 D \psi_2 \, e^{i S_0 [\psi_1, \phi_1] - i S_0 [\psi_2; \phi_2]} 
\label{fpth}
\end{align}
where $\rho_0$ denotes the state (density matrix) of the system, and $U (t_2, t_1; \phi)$ is the evolution operator of the system 
from $t_1$ to $t_2$ in the presence of external sources denoted by $\phi$. %See Fig.~\ref{fig:SK}. 
The sources are taken to be slowly varying functions and there are two copies of them, one for each leg of the CTP contour.
 The second line~\eqref{fpth} is the ``microscopic'' path integral description, with $\psi_{1,2}$  denoting microscopic dynamical variables for the two copies of spacetime of the CTP and $S_0 [\psi; \phi]$ the microscopic action. Whereas in~\eqref{par0} derivatives with respect to sources $\phi$ give thermodynamic quantities of a system, in~\eqref{gen0} derivatives with respect to $\phi_{1,2}$ give dynamical properties of a (non)equilibrium system such as (nonlinear) response and fluctuating functions. 

As in~\eqref{par0} we can consider integrating out short-lived degrees of freedom in~\eqref{fpth} to obtain 
%follow the philosophy of~\eqref{par0} to use a
a non-equilibrium EFT $I_{\rm eff}$ for slow modes (denoting them collectively by $\chi_{1,2}$ and there are now two copies 
of them)
\be
e^{W [\phi_1, \phi_2]}  =  \int D \chi_1 D \chi_2 \, e^{i I_{\rm eff} [\chi_1, \phi_1; \chi_2 , \phi_2; \rho_0]}   \ .
\label{left}
\ee
%with $I_{\rm eff}$ obtained from~\eqref{fpth} by integrating out , %up to some short-distance (time) cutoff $\ep$, 
%and  denoting collectively  remaining slow variables. 
Again to write down the general form of $I_{\rm eff}$  one needs to specify  appropriate dynamical variables  $\chi_{1,2}$ and the symmetries satisfied by the low energy effective action $ I_{\rm eff}$, although these tasks normally become significantly more challenging
in non-equilibrium situations. 
In~\eqref{left} $\rho_0$ is also encoded in the couplings of $I_{\rm EFT}$ (below for notational simplicity we will suppress $\rho_0$ in $I_{\rm eff}$).  In general $I_{\rm eff}$ does not have the factorized form of~\eqref{fpth}, and is complex. 
It is often convenient to introduce the so-called $r-a$ variables~\cite{Chou:1984es,Wang:1998wg} 
\be \label{rava}
\chi_r = \ha (\chi_1 + \chi_2) , \quad \chi_a = \chi_1 - \chi_2, \quad  \phi_r = \ha (\phi_1 + \phi_2) , \quad \phi_a = \phi_1 - \phi_2\ 
\ee
where as usual $\chi_r$ correspond to physical quantities while $\chi_a$ can be interpreted as noises. 

The functional integral~\eqref{left} defines a ``bare'' theory at some short distance (time) cutoff scale.\footnote{The cutoff is chosen so that it is much larger than all microscopic scales, but much smaller than macroscopic scales of questions of interests.}  Physics at larger distance and time scales is obtained by further applying renormalization group procedure.  While $I_{\rm eff}$ in principle
contains  an infinite number of terms with increasingly higher number of derivatives, in practice 
to describe macroscopic phenomena one only needs to keep track of a finite number of relevant interactions.

Non-equilibrium EFTs provide powerful tools for dealing with dynamical questions and
non-equilibrium systems.  The effective action $I_{\rm EFT}$ incorporates dissipations and retardation effects from the bath of short-lived degrees of freedom (which have been integrated out) in a medium. Its general structure has recently been used  to derive from first principle the local second law of thermodynamics~\cite{GL}, and  a new formulation of fluctuating 
hydrodynamics has been proposed in terms of such an EFT~\cite{CGL,CGL1} (see also~\cite{Grozdanov:2013dba,Kovtun:2014hpa,Haehl:2014zda,Harder:2015nxa,Haehl:2015foa,Haehl:2015uoc}). 
%\footnote{For other recent discussions of fluctuating hydrodynamics, see.}. %which systematically incorporate nonlinear interactions involving noises as well as non-equilibrium fluctuation-dissipation relations. 
See also~\cite{Sieberer1} for a review of  applications to driven open systems.
When an $I_{\rm EFT}$ is truncated to quadratic order in noises (i.e. $a$-variables) the path integral~\eqref{left} reduces to a so-called Martin-Siggia-Rose-De Dominicis-Janssen~\cite{msr,Dedo,janssen1} functional integral which is in turn  equivalent to a stochastic Langevin equations (for a review see~\cite{Kamenev}).

Compared to EFTs for equilibrium systems,  there are new elements in identifying both dynamical variables and symmetries for a non-equilibrium EFT~\eqref{left}.  The unitarity of time evolution in~\eqref{gen0} implies that the action should in addition satisfy the following conditions~(see e.g.~\cite{CGL,GL} for more details)
%Taking the complex conjugate of~\eqref{gen0} we find that  $W$ satisfies $W^* [\phi_1, \phi_2] = W[\phi_2,  \phi_1]$ which in turn requires that 
\bega
\label{fer1}
% I^*_{\rm EFT} [\chi_1 , \phi_1; \chi_2, \phi_2] = - I_{\rm eff} [\chi_2 , \phi_2; \chi_1, \phi_1]   \ 
  I^*_{\rm eff} [\chi_r , \phi_r; \chi_a, \phi_a] = - I_{\rm eff} [\chi_r , \phi_r; -\chi_a, -\phi_a]   \\ 
 \label{pos}
{\rm Im} \, I_{\rm eff} \geq 0 \\
 \label{key1}
I_{\rm eff} [\chi, \phi; \chi, \phi] = 0, \quad {\rm or} \quad 
I_{\rm eff} [\chi_r =\chi, \phi_r = \phi; \chi_a=0, \phi_a =0] = 0 \ ,
\end{gather} 
where for definiteness we have taken $\chi_{1,2}$ and sources $\phi_{1,2}$ to be real. 
These conditions are, however, enough only for performing the functional integrals of~\eqref{left} at tree level.  
With loops included one also has to worry about defining the integration measure $D \chi_1 D \chi_2$ precisely. 

To see this, in~\eqref{gen0} taking $\phi_1 = \phi_2 = \phi$, we then find that 
\be \label{nor}
 \Tr \le( U (+\infty, -\infty; \phi) \rho_0 U^\da (+\infty, -\infty; \phi) \ri)  = \Tr (\rho_0) = 1 \quad \Rightarrow \quad 
 W [\phi, \phi] =0   \ .
\ee
While equation~\eqref{key1} leads to~\eqref{nor} at tree-level, this is no longer so when including loops and one has to include an additional integration measure factor. See e.g. Sec. I E of~\cite{CGL} for an explicit discussion.  To ensure~\eqref{nor} at loop level 
can use the standard trick of parameterizing integration measures  by 
introducing an anti-commuting  partner for each bosonic variable, i.e. $c_r, c_a$ for $\chi_r, \chi_a$ respectively, and requiring 
the action to be invariant under the following BRST-type fermionic transformation~\cite{zinnjustin}
\be \label{obrst}
\de \chi_r = \ep c_r, \qquad \de c_a = \ep \chi_a \ .
\ee
Here $\ep$ is an anti-commuting constant. To show that~\eqref{obrst} is enough to ensure~\eqref{nor} at loop level is quite simple 
and is reproduced in Appendix~\ref{app:a} for completeness. In particular, the BRST invariance automatically leads to~\eqref{key1} for the bosonic part of the action. $c_{r,a}$ are anti-commuting but transform the same as their bosonic partners under  spacetime rotations. They will be subsequently referred to as ghost variables following standard terminology. 
%This requirement is natural as anti-commuting degrees of freedom (``ghosts'') and BRST symmetry  provide a convenient way to   and is also anticipated from functional integral form of Langevin equations. 
%With a slight abuse of language, below we will simply refer to such anti-commuting degrees of freedom as fermionic variables. 
%
The need for ghosts and BRST symmetry can also be anticipated  from results on the functional integral forms of stochastic equations~\cite{zinnjustin}, and has been emphasized recently~\cite{CGL,Haehl:2015foa} in the context of fluctuating hydrodynamics.\footnote{See also~\cite{Haehl:2016pec}. Refs~\cite{Haehl:2015foa,Haehl:2016pec} appear to require two BRST generators while~\cite{zinnjustin} and~\cite{CGL} require only one.}

There are three different regimes for~\eqref{left}. The first is the full quantum regime where path integrations describe both quantum and classical statistical fluctuations. The second is the classical regime with $\hbar \to 0$. In the $\hbar \to 0$ limit  the path integrals survive  and describe classical statistical fluctuations. The third is the level of equations of motion which corresponds to 
the thermodynamic limit with all classical and quantum fluctuations neglected. 
Since the constraints~\eqref{fer1}--\eqref{key1} concern only with the general structure of the action, they remain in the classical limit. Similarly the requirement $W [\phi, \phi] =0$ also survives the classical limit, and so do ghost variables and the corresponding BRST symmetry. It is striking that a classical statistical system is significantly constrained by these remnants from quantum unitarity. 
%remain and it is striking that  some ``effective'' anti-commutating degrees of freedom are needed at low energies to ensure unitarity even though the microscopic description may contain only bosonic degrees of freedom.

For many physical processes in nature, macroscopic physical quantities of interests typically vary over spatial and temporal scales much larger than microscopic molecular collision scales (or any microscopic interaction scales).
Such a  system is considered as  in {\it local equilibrium}, for which an additional $Z_2$ symmetry should be imposed on $I_{\rm eff}$~\cite{GL,CGL1}. 

A subclass of local equilibrium systems correspond to thermal systems perturbed by slowly varying external sources, and in this case the need for this $Z_2$ symmetry can be readily understood as follows. For $\rho_0 = {1 \ov Z} e^{-\beta_0 H}$, the generating functional~\eqref{gen0} satisfies an additional constraint coming from combining the Kubo-Martin-Schwinger (KMS) condition~\cite{kubo57,mart59,Kadanoff} with time reversal invariance,
\be
\label{1newfdt1}
 %{\rm KMS \;\; condition:} \quad
W [\phi_1 (x), \phi_2 (x)] = W [\tilde \phi_1 (x); \tilde \phi_2  (x)]
\ee
where $x$ denotes $x^\mu = (x^0, x^i)  = (t, \vx)$ and 
\be
\begin{split} 
\tilde \phi_1 (x) = \phi_1 (-t + i \th, - \vx ) , \qquad 
\tilde  \phi_2 (x) =  \phi_2 (- t - i (\beta_0 - \th), -\vx )  \ . 
\label{tiV}
\end{split}
\ee
for arbitrary $\th \in [0, \beta_0]$. 
Below we will simply refer to~\eqref{1newfdt1} as the KMS condition, but it should be kept in mind it also encodes consequences of microscopic time-reversal symmetry.\footnote{The KMS condition itself only relates $W$ 
to a time-reversed one, and so does the time reversal symmetry of the microscopic theory. Only the combination of them leads to a nontrivial constraint on $W$ itself~\cite{Chou:1984es,Wang:1998wg}. See~\cite{CGL} for a detailed discussion. Depending on circumstances one could combine KMS  with $\sT$ or $\sP \sT$ or $\sC\sP\sT$. For definiteness 
here we follow~\cite{CGL} to combine it with $\sP \sT$. It is simple to adapt~\eqref{tiV} for a system with only $\sT$ invariance by simply removing the minus signs before $\vx$.}   

Additional condition(s) then need to be imposed on $I_{\rm eff}$ for~\eqref{left} to satisfy~\eqref{1newfdt1}. 
  For variables $\chi_{r,a}$  associated with  non-conserved quantities, the required symmetry is well known, probably since 70's~\cite{janssen1,janssen2,Sieberer2}. In this case the couplings between $\chi_{r,a}$ and  external sources 
are the standard ones   
% Turning on external sources $\phi_{1\rmi}, \phi_{2 \rmi}$ for $\chi$'s  we add to~\eqref{w1} 
%the following terms 
\be \label{cex}
\int d^d x \, \le(\chi_{1 \rmi} \phi_{1 \rmi} - \chi_{2 \rmi} \phi_{2 \rmi} \ri) =  \int d^d x \, \le( \chi_{r \rmi} \phi_{a \rmi} + \chi_{a \rmi} \phi_{r \rmi}  \ri)\ 
\ee
which then immediately implies that for~\eqref{left} to satisfy~\eqref{1newfdt1}, the action should satisfy\footnote{One can readily check that the requirements~\eqref{fer1} and~\eqref{dkn} are compatible.} 
\be \label{dkn}
I_{\rm eff} [ \chi_1, \phi_1; \chi_2 , \phi_2] = I_{\rm eff} [ \tilde \chi_1, \tilde \phi_1; \tilde \chi_2 , \tilde \phi_2]
\ee
with 
\be \label{ndym}
\tilde \chi_1 (x) = \chi_1 (-t + i \th, - \vx ) , \qquad 
\tilde  \chi_2 (x) =  \chi_2 (- t - i (\beta_0 - \th), -\vx )  \ . 
\ee
Following~\cite{GL,CGL1} we will refer to~\eqref{dkn} as the dynamical KMS condition. In the absence of external sources 
it becomes a $Z_2$ symmetry (dynamical KMS symmetry) of the action. 
We will refer to  transformations~\eqref{ndym} on dynamical variables as
 dynamical KMS transformations.

The story for variables associated with conserved quantities (hydrodynamical variables) is more complicated since the couplings to external sources are more intricate making it more difficult to deduce the needed transformations on dynamical variables. 
%In 
%due to highly collective nature of the dynamical variables. Two formulations have been proposed so far.
 In~\cite{CGL} a shortcut was proposed which imposes~\eqref{1newfdt1} on a contact-term  action which in turn constrains
 the action for dynamical variables through the special structure of the couplings between dynamical variables and external sources. It was termed as the local KMS condition. % on contact terms of $I_{\rm eff}$ which in turn constrain couplings of dynamical variables.
Only very recently were dynamical KMS transformations on hydrodynamical variables finally found in~\cite{CGL1}.%\footnote{In the classical limit the local and dynamical KMS conditions are equivalent, but both have potential ambiguities at quantum level~\cite{CGL1}.} 

We emphasize that a system in local equilibrium is not restricted to a thermal density matrix in the presence of slowly varying external sources.  Such a system can, for example, be in a pure state.
 In these more general cases while there is no such requirement as~\eqref{dkn}, invariance of an action under dynamical KMS transformations ensures the system is in a local equilibrium.\footnote{In essence, the dynamical KMS condition is  ``local'', i.e. operating at the scale of local inverse temperature, and thus will not care about the global structure of a state, be it a  thermal state or a pure state.}
  For example, in the classical limit dynamical KMS condition ensures that first law and second law of thermodynamics,  as well as fluctuation-dissipation and Onsager relations are all satisfied locally~\cite{GL,CGL1}.  

With dynamical KMS transformations for bosonic variables understood, in this paper we consider the extensions to 
ghost variables, which are needed to have a complete formulation of  a non-equilibrium EFT. 
For example, to ensure~\eqref{1newfdt1} at loop level we need dynamical KMS transformations on all variables. 

Furthermore, it has been long known in the context of functional representation for linear stochastic systems that there is an emergent supersymmetry as a consequence of fluctuation-dissipation relations~\cite{parisi,feigelman,Gozzi:1983rk,Mallick:2010su,zinnjustin}.
  More recently, it was found in~\cite{CGL}  that after imposing the local KMS condition and BRST symmetry there is also an emergent supersymmetry for a  hydrodynamic theory of nonlinear diffusion.  
We would like to understand the precise origin and the full extent of this emergent supersymmetry. 
In particular, we would like to  extend the discussion to a general non-equilibrium EFT including full fluctuating hydrodynamics.\footnote{In~\cite{Haehl:2015foa,Haehl:2015uoc} a certain superalgebra was assumed as a basic input for constructing fluctuating hydrodynamics and an attempt was made to write down the action using superspace. See also~\cite{Haehl:2016pec,Haehl:2016uah}.}

% however was not fully understood. It was not clear whether it happens for full hydrodynamics or other non-equilibrium EFTs. In fact, for the full fluctuating hydrodynamics of~\cite{CGL,CGL1} even imposing the BRST symmetry is technically quite involved and was not attempted there.  It has been conjectured in~\cite{CGL} that the emergent supersymmetry is  important for ensuring~\eqref{1newfdt1} when including loops. 

We will restrict our discussion to the classical level with $\hbar \to 0$. At the classical level, the dynamical KMS transformations dramatically simplify. For example, equations~\eqref{tiV} and \eqref{ndym} become
\bega \label{0mk}
\tilde \phi_{r} (-x) = \phi_{r} (x), \qquad \tilde \phi_{a} (-x) = \phi_{a } (x) + i \beta_0 \p_0 \phi_{r } (x) , \\
\tilde \chi_{r} (-x) = \chi_{r } (x), \qquad \tilde \chi_{a } (-x) = \chi_{a } (x) + i \beta_0 \p_0 \chi_{r } (x)  ,
\label{bkms}
\end{gather}
which are local transformations combined with a spacetime reflection. We stress that these are finite $Z_2$ transformations. 
The dynamical KMS transformations for hydrodynamical variable, although more involved,  have a similar structure (see Sec.~\ref{sec:fluid}). 
The quantum regime has a number of additional complications  and will not be pursued here (see Sec.~\ref{sec:conc} for a brief discussion).

We will show that any system in local equilibrium has an emergent supersymmetry at low energies. 
With increasing complications and generality we consider three classes of systems depending on whether or not a system has conserved quantities or dynamical temperature: (i) no conserved quantities with a fixed background temperature; (ii) with conserved quantities and a fixed background temperature; (iii) with conserved quantities and dynamical temperature.  
Clearly the third class includes all systems. 
As an example for class (i) we consider model A of critical dynamics~\cite{hohenberg}, for class (ii) a theory of nonlinear diffusion,
and for class (iii) a fluctuating hydrodynamics for charged fluids proposed in~\cite{CGL,CGL1}.  It turns out when expressed in terms of the right sets of variables, all three classes have essentially the same structure. 
Here is a summary of the main results:

\ben 

\item  We show that there is essentially a unique extension of  dynamical KMS transformations to ghost variables which is self-consistent. The dynamical KMS transformation on ghost variables turn out to be a $Z_4$ operation, but is still a $Z_2$ operation of the action. 

\item For any action the combination of BRST symmetry and dynamical KMS symmetry leads to an emergent fermionic 
symmetry which together with the BRST symmetry forms a supersymmetric algebra. 

\item Starting with a supersymmetric action one can always construct an action which is both BRST and 
dynamical KMS invariant.

\item Supersymmetry does not impose the full dynamical KMS invariance, only a particular 
consequence of the dynamical KMS symmetry, to which we refer as the special dynamical KMS condition.  Conversely we prove a theorem stating that any bosonic action satisfying the special dynamical KMS condition  is  always supersymmetrizable.

\item For a system for which temperature is non-dynamical (i.e. with a fixed constant temperature), one finds a global supersymmetry. For a system for which temperature is dynamical, such as a fluctuating hydrodynamics, one 
finds a local supersymmetry.

\item Supplementing the bosonic story of fluctuating hydrodynamics proposed in~\cite{CGL,CGL1} with dynamics of ghosts, this paper finally gives a complete formulation of fluctuating hydrodynamics in the classical regime.

\een

The plan of the paper is as follows. In next section we present a general discussion of emergence of supersymmetry 
from BRST and dynamical KMS symmetries. In Sec.~\ref{sec:exm}--\ref{sec:fluid} we discuss three classes of examples.  We 
conclude in Sec.~\ref{sec:conc} with a discussion of future directions. In Appendix A we give further argument for the need of BRST symmetry. Appendix B contains details of a proof for a supersymmetrizability theorem. 

While this paper is in preparation we learned that overlapping results have been obtained by Kristan Jensen,
Natalia Pinzani-Fokeeva, and Amos Yarom~\cite{yarom}.

\section{Emergent supersymmetry: general structure}

In this section we present a general discussion of emergence of supersymmetry from BRST and dynamical KMS symmetries. 
%which we will apply in 
%later sections to specific examples. 

\subsection{General case} 

Consider an action $I[F_i]$  with $F_i = (b_i, f_i)$ which $b_i$ denotes collectively bosonic source and dynamical  fields, 
and $f_i$ denotes collectively anti-commuting source and dynamical fields~(ghost variables here). To make our equations compact we will use  index $i$ to denote both field species and spacetime points. 
We assume that the action is invariant under a BRST-type fermionic symmetry, i.e. 
 \be 
 \de F_i = \ep Q F_i , \qquad  Q F_i {\de I \ov \de F_i} = 0
 \ee 
 where $\ep$ is an anti-commuting constant, and $Q$ is an anti-commuting operator satisfying 
 \be 
 Q^2 F_i = 0, \quad {\rm i.e.} \quad (Q F_j) {\de Q F_i \ov \de F_j} = 0  \ .
 \ee
 Now let us suppose that $I$ is invariant under another bosonic symmetry 
  \be 
 F_i \to  K_\al F_i , \qquad K_\al I [F_i] \equiv I [K_\al F_i] = I [F_i] 
 \ee
 where $K_\al$ is an invertible bosonic operator (i.e. maps bosons to bosons and ghosts to ghosts) and index $\al$ denotes different 
 elements of the symmetry group. 
 Note that while $Q$ acts as a derivation, $K_\al$ acts as a finite transformation. Acting on a product, $K_\al$ transforms all factors 
 at the same time.

 Clearly the action is also invariant under the combined  operations  $Q_\al = K_\al Q K^{-1}_\al$, 
 \be \label{barQ}
 Q_\al F_i = K_\al Q K^{-1}_\al F_i =  \le[(Q F_j) {\p K^{-1}_\al F_i \ov \p F_j} \ri]_{F_i \to K_\al F_i} 
\ee
where the notation on the right hand side means after evaluating $(Q F_j) {\p K_\al^{-1} F_i \ov \p F_j}$ 
replace all $F_i$ by the corresponding $K_\al F_i$. More explicitly 
\be 
0 = K_\al Q  I [F_i] %= Q_\al I [F_i] 
=  K_\al Q  K_\al^{-1} K_\al I [F_i ] = Q_\al I [F_i] \ .
%Q I [ K^{-1}_\al F_i] = K \le[Q F_j {\de \tilde F_i \ov \de F_j} {\de I [\tilde F_i] \ov \de \tilde F_i} \ri] 
%=\bar Q F_i {\de I [F_i] \ov \de F_i}
\ee
%where in the above equation $\tilde F_i = K^{-1} F_i$. 
By definition 
\be 
 Q^2_\al = 0 \ . 
\ee
Thus we find that for each symmetry transformation $K_\al$ there is an emergent fermionic symmetry $Q_\al$. Note that the collection $\{Q_\al \}$ also includes the original $Q$ as $\{K_\al\}$ includes the identity element. 
Note that %The algebra among $\{Q_\al \}$ is then given by 
\be \label{anc}
\{Q_\al , Q_\beta\} =   Q_\al Q_\beta + Q_\beta  Q_\al = K_\al \{Q_{\al^{-1} \beta}, Q\} K_\al^{-1}, \quad 
K_{\al^{-1} \beta} \equiv K_\al^{-1} K_\beta  %Q K Q K^{-1} + K Q K^{-1} Q 
 \ .
\ee

Suppose we have an action $I_0 [F_i]$ which is {\it not}  invariant  under a $K_\al$-transformation. 
Then it follows immediately from our definition that 
\be \label{QI1}
Q I_0 [F]=0\quad \iff  \quad Q_\al {I}_\al [F]=0 \ 
\ee
where 
\be \label{i1}
I_\al [F_i] \equiv K_\al I_0 [F_i] = I_0 [K_\al F_i]  \ .
\ee 

\subsection{A special case}

Now let us specialize to a situation which will be relevant for the rest of this paper, with $\{K_\al\} = {1, K, K^2, K^{-1}}$
being a set of $Z_4$ transformations satisfying 
\be \label{k0}
K^2 b_i = b_i, \qquad K^2 f_i = - f_i \ .
\ee
In this case for any action $I_0$ (which is not necessarily invariant under $K$) we have 
\be \label{i2} 
K^2 I_0 [F_i] = I_0 [F_i]
\ee
as an action is always even in the number of ghosts variables. Also note that 
$Q_{K^2} = - Q$ and $Q_{K^{-1}} = - Q_K$, and thus the independent $\{Q_\al\}$'s are
$Q$ and  $\bar Q \equiv Q_K$. %We will see later from the explicit form of dynamical KMS transformations 
%the anti-commutator of $Q, \bar Q$ in fact forms a supersymmetric algebra. 

From~\eqref{i1} and~\eqref{i2} we then have  for any action $I_0$
\be \label{z22}
Q \tilde I_0 [F_i] = 0 \quad \iff   \quad \bar{Q} I_0 [F_i]=0 
\ee
where 
\be 
\tilde I_0 [F_i] \equiv K I_0 [F_i] = K^{-1} I_0 [F_i] \ .
\ee

%Now let us suppose $K$ is a $Z_2$ operation of $I_0$, but it is not necessarily a $Z_2$ operation on $F_i$), i.e.  
%\be \label{z21}
%K^2 I_0 [F_i] = I_0 [K^2 F_i] = I_0 [F_i], \quad \to \quad K I_0 [F_i] =  K^{-1} I_0 [F_i] \ .
%\ee 
%In this case we then also have 
Now suppose $I_0$ is BRST invariant, i.e. $Q I_0 =0$. 
We can construct a $K$-invariant action as 
\be \label{susm}
I = \ha (I_0 + \tilde I_0)  \ .
\ee
But this action is in general not BRST invariant as $Q \tilde I_0$ does not have to be zero. From~\eqref{z22} we conclude that 
for $I$ to be both BRST and $K$-invariant, the sufficient and necessary condition is that $I_0$ should in addition be invariant under $\bar Q$. 
%Of course from our earlier discussion $I$ is also invariant under $\bar Q$. 

%Thus we find that  if $I_0$ is invariant under both $Q$ and $\bar Q$, then we can always construct an action which is both BRST and  $K$-invariant.  

 \subsection{Strategy for extending dynamical KMS transformations to ghosts}
 
 Since ghost variables are introduced to give the correct integration measure and do not directly couple to external sources, there is no obvious principle to determine how they should transform under dynamical KMS symmetry. Our strategy is based on the following non-trivial self-consistency requirement: BRST and dynamical KMS invariance of the full action does not put further constraint on the pure bosonic part of the action. More explicitly, with the full action written in a form 
\be 
I [b_i, f_i] = I_b [b_i] + I_f [b_i, f_i]
\ee
then the pure bosonic part $I_b[b_i]$ should coincide with the most general action one can construct based~\eqref{fer1}--\eqref{key1} and the bosonic dynamical KMS invariance. This requirement is due to that the bosonic action $I_b$ already provides a complete formulation of tree-level physics, thus extension of dynamical KMS symmetry to the ghost sector should not change that physics. The requirement is highly nontrivial mathematically as dynamical KMS invariance constrains the ghost part of the action $I_f$ which in turn constrains the bosonic part $I_b$ via BRST symmetry. 
 
Our discussion contains the following elements: 
 
 \ben 
 
 \item Applying the consistency requirement at quadratic level in dynamical variables uniquely determines the 
 dynamical KMS transformation for ghost variables at linear level. 
 
% there is a unique action which is BRST invariant. 
%Thus should be a symmetry of that action, which determines the transformation at linear level. In particular, being linear, the transformation only involves ghost variables themselves. \textcolor{red}{(PG: There is a small subtlety I mentioned below around 3.6)}

\item As a simplest possibility we postulate the linear transformation deduced from the quadratic action is the full transformation. 
Including both bosons and ghosts, the dynamical KMS transformations  have the structure discussed around~\eqref{k0}. 
We then construct $\bar Q$ explicitly  from $Q$ and $K$ using~\eqref{barQ}. 
One finds that $Q, \bar Q$ form a supersymmetric algebra. 
In other words, {\it a non-equilibrium EFT must be supersymmetric invariant}.

\item We provide a strong support for the postulate of item 2 by proving that the self-consistency requirement is indeed satisfied for the full nonlinear action. %In other words, applying the linear transformation to full non-linear action still satisfies 
%a rather non-trivial self-consistency requirement: BRST and dynamical KMS invariance of the full nonlinear action does not put further constraint on the pure bosonic part of the total action other than 
%the bosonic dynamical KMS invariance. This is due to that the bosonic action already provides a complete formulation of tree-level physics, thus extension of dynamical KMS symmetry to the ghost sector should not change that physics. The requirement is highly nontrivial mathematically as dynamical KMS invariance constrains the ghost part of the action which is in turn related to the bosonic part 
%via BRST symmetry. 
%Modifying the dynamical transformation on ghosts by adding nonlinear terms should spoil the self-consistency. 
 
\een
From the above discussion and that around~\eqref{susm} we conclude that one can obtain a BRST and dynamical KMS invariant action by first writing down a most general supersymmetric action and then impose~\eqref{susm}.

\subsection{Special dynamical KMS symmetry and a theorem on supersymmtrizability}  \label{sec:B4}

In this subsection we elaborate a bit further on the self-consistency requirement of the previous subsection. 

Consider a most general action $I_b [b_i]$ of bosonic variables which satisfies~\eqref{fer1}--\eqref{key1} and the dynamical KMS condition~\eqref{dkn}. Now adding a ghost partner $f_i$ for each bosonic variable
to obtain a full action $I [b_i, f_i]$ which is BRST and dynamical KMS invariant. From our discussion above we learned that 
this full action must be supersymmetric. The self-consistency requirement requires that the bosonic part of $I[b_i, f_i]$ should coincide with the original $I_b$. This in turn requires that $I_b$ be supersymmetrizable. 
Conversely, if $I_b$ is supersymmetrizable, then we can construct a BRST and dynamical KMS invariant action with the same bosonic part by first constructing a supersymmetric extension of $I_b$ and then using~\eqref{susm}. Thus for $I_b$ to be supersymmetriable is both sufficient and necessary for constructing a full BRST and dynamical KMS invariant action. 
So the self-consistent condition boils down to the statement:  a bosonic action which satisfies~\eqref{fer1}--\eqref{key1} and the bosonic dynamical KMS condition~\eqref{dkn} should be supersymmetrizable under the supersymmetry generated by $Q$ and $\bar Q$. 

We will be able to prove that this is indeed the case. In fact we will be able to prove a stronger statement which was first 
observed in~\cite{CGL} for a theory of nonlinear diffusion at cubic level. To describe the statement, we need to be a bit more specific on the general structure of dynamical KMS condition. %~\cite{GL,CGL1}. 

Equations~\eqref{fer1} and~\eqref{key1} imply that the bosonic Lagrangian density $\sL_b$ can be expanded in $a$-fields as 
\be \label{w3}
\sL_b = \sum_{n=1}^\infty \sL_b^{(n)} = \sum_{n=1}^\infty i^{\eta_n}   f^{(n)} [\Lam_r]  \Phi_a^{n},  \qquad \eta_n = \bca 1 & n\; {\rm even} \cr
             0 & n \; {\rm odd} 
             \eca \ 
\ee
where we use $\Lam_r, \Phi_a$ to denote collectively $r$- and $a$-fields respectively. Note that the sum starts with $n=1$ 
as $n=0$ term is not allowed by~\eqref{key1}.  In the classical limit $\hbar \to 0$, the dynamical KMS transformation on bosonic variables is a $Z_2$ transformation which can be schematically written as\footnote{Clearly~\eqref{0mk}--\eqref{bkms} are the of the form~\eqref{glkms}. Those for hydrodynamical variables are given explicitly in Sec.~\ref{sec:diff} and Sec.~\ref{sec:fluid}.}
\be  \label{glkms}
\tilde \Lam_{r} (-x) = \Lam_{r} (x),  \qquad  \tilde \Phi_{a} (-x) = \Phi_{a} (x) + i \Phi_{r} (x)
\ee
where $\Phi_r$ is a product of bosonic $r$-variables with altogether one derivative. The dynamical KMS condition~\eqref{dkn} can then be written as 
\be \label{ryt}
\tilde \sL _b =  \sL_b + \p_\mu V^\mu 
\ee
where $\tilde \sL_b$ is obtained by plugging~\eqref{glkms} into~\eqref{w3} and taking $x \to - x$, i.e. 
\be \label{ryt1}
\tilde \sL_b  = \sum_{n=1}^\infty \widetilde{\sL_b^{(n)}} = \sum_{n=1}^\infty i^{\eta_n}  f^{(n) *} [\Lam_r] (\Phi_a + i \Phi_r)^n  = \sum_{k=0}^\infty \le(\tilde \sL_b \ri)_k  \ .
\ee
$f^{(n)*}$ is obtained from $f^{(n)}$ by flipping the signs of all derivatives and $\le(\tilde \sL_b \ri)_k$ denotes  terms in $\tilde \sL_b$ with $k$ factors of $\Phi_a$. Note that the $k$-sum starts with zero. Equating equation~\eqref{ryt} order by order in the expansion of $\Phi_a$ we then find an infinite number of conditions
\be \label{skm1}
\le(\tilde \sL_b \ri)_0 = \p_\mu V^\mu_0 
\ee
and 
\be\label{gkmd}
\le(\tilde \sL_b \ri)_k=  \sL_b^{(k)} + \p_\mu V^\mu_k, \qquad k \geq 1
\ee  
where $V_k^\mu$ denotes terms containing $k$ factors of $\Phi_a$.

Alternatively we can also impose the dynamical KMS condition as follows. 
Take a Lagrangian density $\sL_0$ of the form~\eqref{w3}. Due to $Z_2$ nature of the transformation, then
\be \label{w4}
\sL_b =\ha \le( \sL_0 + \tilde \sL_0 \ri),
\ee
automatically satisfies~\eqref{dkn}.  But as in~\eqref{ryt1} $\tilde \sL_0$ contains terms with no $\Phi_a$, and we must  require that such terms in $\tilde \sL_0$ vanish, which is precisely~\eqref{skm1}. Thus it is enough to impose~\eqref{skm1} and~\eqref{w4} as all the conditions~\eqref{gkmd} with $k \geq 1$ are automatically taken care of by~\eqref{w4}. 
We will refer to~\eqref{skm1} as the special dynamical KMS condition. 

From~\eqref{z22}--\eqref{susm} and that BRST symmetry implies~\eqref{key1} for the bosonic part, it then follows that supersymmetry of $\sL_0$ ensures the special dynamical KMS condition for its bosonic part.
Conversely, we will be able to prove the following supersymmetrizability theorem: 

{\it Any local bosonic Lagrangian which satisfies~\eqref{key1} and the special dynamical KMS condition~\eqref{skm1}
is supersymmetrizable.} 

Comparing with the discussion around~\eqref{susm}, we see the procedure of imposing~\eqref{w4} commutes with supersymmetrization. One could either do it before or after.

\section{With no conserved quantities at a fixed temperature: model A}  \label{sec:exm}

In this and next two sections we consider the extension of dynamical KMS transformations to ghosts
and the associated supersymmetry for some explicit examples of non-equilibrium EFTs. 
In this section we will consider systems with no conserved quantities at a fixed background temperature, i.e. temperature is not a dynamical variable. In this case the story is technically much simpler, but captures all the essential elements.

%\subsection{Model A}  \label{sec:moda}

As an illustration of a system with no conservation laws, we consider the critical dynamics of a $n$-component real order parameter $\chi_\rmi, \rmi=1,\cdots , N$ at a fixed inverse temperature $\beta_0$~(i.e. model A~\cite{hohenberg,Folk}). 
The dynamical variables in~\eqref{left} are then $\{\chi_{r\rmi}, \chi_{a\rmi}\}$ and the action should be invariant under
an $SO(N)$ symmetry which rotates  $\chi_{r\rmi}, \chi_{a\rmi}$ simultaneously\footnote{The boundary condition for CTP requires that $\chi_1 = \chi_2$ at $t=\infty$, thus any global symmetry must rotate $\chi_{1,2}$ together.}. 
The dynamical KMS transformation for bosonic variables is the same as~\eqref{bkms} 
\bega \label{ckms1}
\tilde \chi_{r \rmi} (x)  = \chi_{r  \rmi}(-x) , \qquad \tilde \chi_{a \rmi} (- x) = \chi_{a \rmi} (x) +{ i \beta_0  } \p_0 \chi_{r \rmi} (x)   \  .  %\\
%\tilde \phi_r (k) = \phi_r (-k) , \qquad  \tilde \phi_a (k) = \phi_a (-k) - \beta \om  \phi_r (-k)  \ .
\end{gather} 
In this case the couplings~\eqref{cex} to external sources are rather trivial, so we will suppress the sources below.  
At quadratic order in $\chi_{a,r}$ (but to all orders in derivatives) the bosonic part of the Lagrangian can be written as 
\be 
\sL_b = \chi_{a \rmi} G_{ra} \chi_{r \rmi}  + {i \ov 2} \chi_{a \rmi} G_{aa} \chi_{a \rmi}
\ee
where $G_{ra}$ are $G_{aa}$ are some differential operators. Note that by definition $G_{aa}$ satisfies $G_{aa} = G^*_{aa}$ where $G^*_{aa}$ denotes the operator 
obtained from $G_{aa}$ by taking all $\p_\mu$ to $- \p_\mu$. 
Imposing the dynamical KMS condition leads to the condition 
\be \label{skms}
G_{ra} - G^*_{ra} = - \beta_0 \p_0 G_{aa}  \ .
\ee

As discussed in the Introduction we should also introduce anti-commuting partners $c_{r \rmi} , c_{a \rmi}$
for $\chi_{r \rmi},\chi_{a \rmi}$ respectively, and require the action to be invariant under the following BRST transformations
\be \label{brst0}
\de \chi_{r \rmi} \equiv \ep Q \chi_{r \rmi} = \ep c_{r \rmi}, \quad \de c_{a \rmi} \equiv \ep Q c_{a \rmi}  = \ep \chi_{a \rmi}, \quad
Q \chi_{a \rmi} = Q c_{r \rmi} =0  \ .
\ee
At quadratic level the most general Lagrangian invariant under~\eqref{brst0}  can be written as\footnote{There cannot be a $c_a K_a c_a$ term as it is incompatible with BRST symmetry.} 
\be \label{qa0}
\sL =  \chi_{a \rmi} G_{ra}  \chi_{r \rmi}  + {i \ov 2} \chi_{a \rmi} G_{aa} \chi_{a \rmi}  - c_{a \rmi} G_{ra}  c_{r \rmi} 
+ c_{r \rmi} G_{rr} c_{r \rmi}  \ 
\ee
where $G_{rr}$ is an arbitrary differential operator satisfying $G_{rr} = - G_{rr}^*$ from anti-commuting nature of $c_{r \rmi}$.
At quadratic level the dynamical transformation on $c_{r,a}$ must be linear and requiring no further constraints on $G_{ra}$ 
we find that  the only possibility is to require Lagrangian be invariant under\footnote{One can in fact consider $c_a \ra \a c_r$ and $c_r \ra -\frac{1}{\a} c_a$ for any real $\a$, but such an $\al$ can be absorbed by redefining $c_a$.}
\be \label{fkms}
c_{a \rmi} \to \tilde c_{a \rmi} (x) = c_{r \rmi} (-x), \qquad c_{r \rmi}  \to \tilde c_{r \rmi} (x)= - c_{a \rmi} (-x) \ 
\ee
which in turn requires $G_{rr} =0$. We thus propose~\eqref{fkms} as the dynamical KMS transformation for ghosts. %With~\eqref{skms} then the Lagrangian 
%is invariant under both BRST transformations and dynamical KMS transformations~\eqref{ckms1}, ~\eqref{fkms}

Combining~\eqref{ckms1} and~\eqref{fkms} we find the $Z_4$ structure discussed around equation~\eqref{k0}
with 
\bega
\label{k11}
K \chi_{r \rmi} (x) =  \chi_{r \rmi} (-x), \qquad  K \chi_{a \rmi} (x) =  \chi_{a \rmi} (-x) + i \beta_0 \p_0 \chi_{r \rmi} (-x), \\
K c_{a \rmi} (x)= c_{r \rmi} (-x), \qquad  K c_{r \rmi} (x)= - c_{a \rmi} (-x) \ .
\label{k22}
\end{gather} 
Now applying~\eqref{barQ} to~\eqref{k11}--\eqref{k22} and~\eqref{brst0},  we find that 
\be \label{susy0}
\bar \de \chi_{r \rmi} \equiv  \bar \ep \bar Q \chi_{r \rmi} =  - \bar \ep c_{a \rmi}, \quad \bar \de \chi_{a \rmi}  =  i \bar \ep  \beta_0 \p_0 c_{a \rmi}  , \quad
\bar \de c_{r \rmi} =   \bar \ep \le(\chi_{a \rmi}  +{ i \beta_0  } \p_0 \chi_{r \rmi} \ri) , \quad
\bar \de c_{a \rmi} = 0 \ .
\ee
It can also be readily checked that 
\be \label{susya}
\{Q , \bar Q \}  = i \beta_0 \p_0  \
\ee
i.e. $Q, \bar Q$ form a supersymmetric algebra. It can also be checked explicitly that invariance of~\eqref{qa0} under $Q$ and $\bar Q$ indeed leads to~\eqref{skms}.

%Now at full nonlinear level we require the action to be supersymmetric invariant and then the dynamical KMS invariant action can be obtained by applying~\eqref{susm}. Note that equation~\eqref{fer1} can be generalized straightforwardly to include ghosts. 

\subsection{Superspace} 

To impose supersymmetry, 
%To prove the theorem discussed below~\eqref{w4} at the end of Sec.~\ref{sec:B4}, 
it is convenient to 
use superspace formalism~\cite{Bagger1992}. 
%and to write down general supersymmetric theories.
%Here the supersymmetric algebra~\eqref{susya} is quite small. 
We introduce two Grassmannian coordinates ${\th, \bar{\th}}$ and  the superfield
\be\label{sf0}
\Psi_{\rmi} = \chi_{r \rmi}+ \th c_{r \rmi} +c_{a \rmi}  \bar \th   + \th  \bar \th   \chi_{a \rmi}  \ .
\ee 
$Q, \bar Q$ can then be written in terms of the following differential operators
\begin{equation}
Q=\p_{\th},\quad\bar{Q}=\p_{\bar{\th}} - i \th \beta_0  \p _0
\end{equation}
with~\eqref{brst0} and~\eqref{susy0} given by  
\be
\de \Psi_\rmi=(\ep Q+\bar{\ep}\bar{Q}) \Psi_\rmi \ .
\ee
Note that as usual  acting on superfields 
\be 
\{Q , \bar  Q\} = - i \beta_0  \p _0
\ee
with an opposite sign from~\eqref{susya}. 

The corresponding covariant derivatives are
\be
\bar{D}=\p_{\bar{\th}},\quad D=\p_{\th} + i\beta_0 \bar{\th} \p_0
\ee
which satisfy
\begin{equation}
D^{2}=\bar{D}^{2}=0,\quad\{D,\bar{D}\}= i\beta_0\p_{0},\quad\{Q,D\}=\{Q,\bar{D}\}=\{\bar{Q},D\}=\{\bar{Q},\bar{D}\}=0 \ .
\end{equation}
Note that
\begin{align} \label{sfield}
\bar{D}\Psi_\rmi &=-c_{a\rmi}- \th \chi_{a\rmi},\quad &D\Psi_\rmi &=c_{r\rmi}+ \bar{\th}\tilde \chi_{a \rmi} (-x) -i\beta_0 \t\bar{\t} \p_{0}c_{r\rmi}\\
\bar{D}D\Psi_\rmi  &= \tilde{\chi}_{a\rmi} (-x)+ i\beta_0 \th\p_0 c_{r\rmi},\quad &D\bar{D}\Psi_\rmi &=-\chi_{a\rmi} - i\beta_0 \bar{\th}\p_{0}c_{a\rmi} + i\beta_0 \t \bar{\t} \p_0 \chi_{a\rmi}
\label{sfi1}
\end{align}

A general Lagrangian which is invariant under~\eqref{brst0} and~\eqref{susy0} can then be written as 
\be 
\sL = \int d \bar \th d \th \, \sF [\Psi_\rmi, D, \bar D , \p_\mu] 
\ee
where $\sF$ is a local expression constructed out of $\Psi_{\rmi}$, and their covariant and ordinary derivatives. 

\subsection{Proof of the supersymmetrizability theorem}  \label{sec:the}

We now present a proof of the supersymmetrizability theorem stated at the end of Sec.~\ref{sec:B4}. 
Here we will discuss the main steps. There is a key step whose proof is rather contrived, which we will leave to Appendix~\ref{app:proof}. 

Consider a general bosonic action $I_b [\chi_{r \rmi},\chi_{a\rmi}]$, which satisfies~\eqref{fer1},~\eqref{key1}, and the special dynamical KMS condition~\eqref{skm1}.  Since the dynamical KMS transformation~\eqref{ckms1} is linear in fields, it does not change the total number of fields in a given term. In other words, suppose we expand $\sL_b$ in terms of the power of dynamical variables 
\be 
\sL_b = \sum_{n=2}^\infty \sL_n
\ee
where $\sL_n$ contains altogether $n$ factors of $\chi_{r, a}$, then different $\sL_n$'s  do not mix under dynamical KMS transformations. It is then enough to prove the theorem for a general $\sL_n$.

$\sL_n$ can be written schematically in a form 
\begin{equation} \label{lb}
\sL_{b}=\sum_{m=1}^n i^{\eta_m} f^{(m,n-m)} \chi_a^m \chi_r^{n-m} 
\ee 
where each term should be understood as
\be 
 f^{(m,k)} \chi_a^m \chi_r^{k}  =  f^{(m,k)}_{I_1 \cdots I_m J_1 \cdots J_{k}} \chi_a^{I_1} \cdots  \chi_a^{I_m}
  \chi_r^{J_1} \cdots  \chi_r^{J_k}
 \ee
and the indices $I, J$ include both species indices and indices for all possible derivatives on them. It is easy to write them 
in momentum space, for example, 
\be 
a_{\rmi \rmj \rmk} (k_1, k_2, k_3)  \chi_{a \rmi} (k_1)   \chi_{r \rmj} (k_2)  \chi_{r \rmk} (k_3)  %+ a_{\rmi \rmj, \mu \nu \lam} \p_\mu \p_\nu \chi_{a \rmi} \p_\lam \chi_{r \rmj} + \cdots 
\equiv a_{IJ_1 J_2} \chi^I_a \chi_r^{J_1}  \chi_r^{J_2}
\ee
with $I = (\rmi, k_1^\mu)$ and similarly for $J_1, J_2$.  $f^{(m,k)}$ is then symmetric among the first $m$ and last $k$ indices. 

Now take any term in~\eqref{lb} with $m > 1$,  choose a $\chi_a$ factor and replace it by $\tilde \chi_a  (-x) - i \beta_0 \p_0 \chi_r $. The first term resulted from the replacement has the form 
\be \label{f1}
f^{(m,n-m)} \chi_a^{m-1} \tilde \chi_a (-x)  \chi_r^{n-m}
\ee
which can be supersymmetrized as 
\be \label{f2}
\int d \bar \th d \th \, f^{(m,n-m)} (-\bar D \Psi) (-D \bar D \Psi)^{m-2}  D \Psi \Psi^{n-m} 
\ee
where we have used~\eqref{sfield}--\eqref{sfi1}. Note that in~\eqref{f2} the only 
 pure bosonic term is~\eqref{f1}.  
 The second term resulted from the replacement can be regrouped into terms with $m-1$ $\chi_a$'s. Continuing this procedure we will then be left with terms with one factor $\chi_a$ which we will denote as 
\be \label{ibb}
\sI_b = g_{J_1 \cdots J_{n}}  \chi_a^{J_1}  \chi_r^{J_2} \cdots \chi_r^{J_{n}}, %\qquad k = n-1
\ee
where $g$ is symmetric in $J_2, \cdots J_n$ indices. 

Note that under a dynamical KMS transformation, a term of the form~\eqref{f1} will always contain at least one factor of $\chi_a$ due to the $\tilde \chi_a$ factor there.  Thus the special dynamical KMS condition~\eqref{skm1} will only involve~\eqref{ibb} which  can be written as 
\be \label{skm2}
i \beta_0 g_{J_1 \cdots J_n}  \p_0 \chi_r^{J_1}   \chi_r^{J_2} \cdots \chi_r^{J_{n}} = \p_\mu V^\mu_0
\ee  
and in momentum space 
\be \label{skm3}
\om_1  g_{ J_1 \cdots J_n} \chi_r^{J_1} \chi_r^{J_2}   \cdots \chi_r^{J_{n}} = 0   \ .
\ee
Recall that index $J_1 \cdots$ include both species indices and momenta, i.e. $\chi_r^{J_k} \equiv \chi_{r \rmi} (\om_k, \vk_k)$
 with $k=1, \cdots n$, and momentum conservation implies that $\sum_{k=1}^n \om_k =0$. 
%\be 
%\om + \sum_{k=1}^{n-1} \om_k = 0 \ . 
%\ee

From properties of symmetric polynomials one can show that~\eqref{skm2} implies that 
$\sI_b$ can be written as (the proof of which is a bit involved and we leave it to Appendix~\ref{app:proof})
\be \label{dfe}
\sI_b = \sI_b^{(s)} + \sI_b^{(a)}
\ee
where
\ben 

\item in $\sI_b^{(s)}$ the corresponding  $g_{J_1 \cdots J_{n}}$  is fully symmetric under exchanges of its indices, for which using~\eqref{sf0} $\sI_b^{(s)}$ can be supersymmetrized as 
\be 
\int d \bar \th d \th  \, g_{J_1 \cdots J_{n}}\Psi^{J_1} \cdots \Psi^{J_{n}} \ .
\ee

\item $\sI_b^{(a)}$ can be written in a form 
\be \label{uei}
\sI_b^{(a)} = h_{IJ} (\chi_r)  \chi_{a}^{ I} \p_0 \chi_{r}^{J} , \qquad h_{IJ} = - h_{JI} \ .
\ee  
Using~\eqref{sfield}--\eqref{sfi1}
such a term can be supersymmetrized as 
\be 
i\b_0^{-1}\int d \bar \th d \th \, h^{IJ} (\Psi) \bar D \Psi_I D \Psi_J  \ .
\ee

\een
We thus have shown all terms in $\sL_n$ can be supersymmetrized,  which concludes the proof. 
To conclude this subsection let us note that
the proof does not depend on the nature of the species index $\rmi$, which can be generalized to any kinds of indices 
including spacetime indices, say for a tensor field. In particular we will see the proof applies also to the examples of next two sections. 

%\item $\chi_I$ for different values of index $I$ do not have to be independent. 

%\item In addition to fields which transform under supersymmetric transformations the Lagrangian can depend 
%on any other fields which do not transform under supersymmetry. In particular there is no need for the dependence on these 
%fields to be polynomial. 

%\een

\subsection{Full formulation}

To complete the formulation  of the model A EFT, now let us consider the generalization of~\eqref{fer1} to the full action. 
A natural generalization is 
\be
  I^*_{\rm eff} [\chi_r , c_r ; \chi_a,  c_a] = - I_{\rm eff} [\chi_r , \eta_r  c_r; - \chi_a, \eta_a c_a]   
\ee
with $\eta_{r,a} = \pm 1$. It can be readily checked that only the choice $\eta_r = 1$ and $\eta_a =-1$ is compatible with the BRST symmetry, and thus we should have 
\be \label{fer2}
  I^*_{\rm eff} [\chi_r , c_r ; - \chi_a,  - c_a] = - I_{\rm eff} [\chi_r ,  c_r;  \chi_a,  c_a]    \ . 
\ee
To see that~\eqref{fer2} is also compatible with $\bar Q$ and $K$ operations,  let us define an operation $\hat S$ as 
\be 
\hat S F_r \equiv F_r^*, \qquad \hat S F_a \equiv - F_a^*, \qquad \hat S I [F_r, F_a] \equiv I^*[F_r, - F_a] 
\ee 
where $F_{r,a}$ denote respectively any $r$ and $a$-type variables (including sources, bosonic and ghost dynamical variables). 
Equation~\eqref{fer2} can then be written as
\be 
\hat S I_{\rm eff} = - I_{\rm eff}
\ee
Now it can be readily checked that 
\be 
[Q,\hat S]= 0, \quad \{\bar{Q},\hat S\}=0 
\ee
and thus supersymmetry is preserved by $\hat S$. Also note that 
$[K,\hat S]=0$ acting on bosonic fields and $\{K, \hat S\}=0$ acting on ghost fields, and thus acting on action $K$ commutes with $S$ due to the fact that an action must contain even number of ghost fields. This shows~\eqref{fer2} is also compatible with 
dynamical KMS condition.  
 
We also need to check the self-consistency:~\eqref{fer2} should not put further constraints on the bosonic action.  
This amounts to showing the full action $I$ obtained by  supersymmetrizing a bosonic action satisfying~\eqref{fer1}  satisfies $\hat S I = - I$.  Note that from the discussion of last section any term in the bosonic action can be supersymmetrized to a single term in terms of superfields. We thus only need to show that any term in the superspace has a definite eigenvalue under $\hat S$ (then this eigenvalue must agree with that of the bosonic part).  To see this note that the superfield~\eqref{sf0} has the following structure under transformation of $\hat S$:  $\Psi\sim X+i\bar{\t}X$, $\bar{D}\Psi\sim iX$, $D\Psi \sim X+i\bar{\t}X$, $\bar{D}D\Psi\sim iX$ and $D\bar{D}\Psi\sim iX+\bar{\t}X$ where $X$ denotes the type of fields with eigenvalue $1$ under $\hat S$. Note that $X$ does not contain $\bar{\t}$. Thus any term consisting of products of such superfields will be the proportional to $X^n(X+i\bar{\t}X)^m\sim i\bar{\t}X^{n+m}$, where only one $\bar{\t}$ survives. It has a definite eigenvalue of $\hat S$. 
Finally given that $[K, \hat S] =0$ acting on action, the step~\eqref{susm} does not change the eigenvalue of $\hat S$. 

We can now present the full procedure for constructing the EFT for model A using supersymmetry: 

\ben

\item Construct a most general supersymmetric action $I_0$, which satisfies~\eqref{fer2} (and of course whatever other symmetries of the system). 

\item Construct the full action using~\eqref{susm}. 

\item The bosonic part of the action should further be constrained by~\eqref{pos}. 

\een
Instead of using supersymmetry one can of course directly impose BRST symmetry and the special dynamical KMS condition.  
With the powerful formalism of superspace, supersymmetry should in general  be a faster route.  

 Finally we note that in a most general supersymmetric action there can be terms which are not related to the pure bosonic action, i.e. terms involving ghosts transform among themselves under supersymmetric transformations. Whether one should include such terms requires further consideration.

\section{With conserved quantities at a fixed temperature: nonlinear diffusion }\label{sec:diff}

In this section as an example of systems with conserved quantities at a fixed temperature we consider the hydrodynamic theory for nonlinear diffusion developed in~\cite{CGL}. For slow variables associated with conserved quantities, couplings to external sources play an important role in the formulation of the theory. So in this section we will turn on external sources from the beginning. 
We will see that the same structure as that of model A emerges. % regarding  dynamical KMS transformations for ghosts, and supersymmetry is rather similar to.  
%but there is a richer structure regarding external sources. 
The discussion here generalizes and systemizes some previous observations in~\cite{CGL} regarding BRST invariance, 
KMS conditions and supersymmetry.

We consider the theory of diffusion mode associated  with a $U(1)$ conserved current at a fixed inverse temperature $\beta_0$,   ignoring possible couplings between the diffusion mode and other hydrodynamical modes.
The dynamical variables are $\vp_r, \vp_a$ with $\vp_r$ interpreted as the diffusion mode and $\vp_a$ the corresponding noise variable. The background sources are $A_{r \mu}$ and $A_{a \mu}$ which couple to conserved currents $J_{a}^{ \mu}$ and $J_{r}^{\mu}$ respectively. The bosonic action satisfy the following conditions: 
 
 \ben
 
 \item  $\vp_r, \vp_a$ must always be acted on by  at least one derivatives. We will thus count 
  $\p_\mu \vp_{r,a}$  as having zeroth derivative. In the presence of background fields $A_{r\mu}, A_{a \mu }$, 
  the action should depend only on  the combinations 
 \be \label{bde}
 B_{r \mu} = A_{r \mu} + \p_\mu \vp_r, \qquad B_{a \mu} = A_{a \mu} + \p_\mu \vp_a \ 
 \ee
 i.e. 
 \be 
 I_b [\vp_r, A_{r \mu}; \vp_a , A_{a \mu}] = I_b [B_{r \mu}, B_{a \mu}] \ .
 \ee
 The local chemical potential is given by $\mu = B_{r 0} = A_{r 0} + \mu_d $ with $\mu_d = \p_0 \vp_r$ giving the dynamical part, and it is often convenient to use 
 $\hmu = \beta_0 \mu$ and $\hat \mu_d = \beta_0 \mu_d$. 
 
 \item The action is invariant under
 \be \label{csh}
\vp_r \to \vp_r - \lam (\sig^i) , \qquad \vp_a \to \vp_a   \ .
\ee

\een
The dynamical KMS transformation on bosonic variables  are 
\be
\tilde \vp_r (x) = - \vp_r (-x), \qquad
\tilde \vp_a (x) = - \vp_a (-x)-i \beta_0 \p_0 \vp_r (-x)
% i \hat \mu_d (- x) \
 \ee
 and when including background fields 
\bega %\label{ckms2}
%\tilde \mu_d (-x) = \mu_d (x), \qquad \p_\mu \tilde \vp_a (-x) = \p_\mu \vp_a (x) + i \p_\mu \hmu_d (x)  \\
 \label{ckmsn}
\tilde B_{r\mu} (-x)  = B_{r \mu}  (x) , \quad \tilde B_{a \mu} (-x) = B_{a \mu} (x) +i  \beta_0 \p_0 B_{r \mu}  \ .
% \Phi_{r \mu}  (x) ,  \quad \Phi_{r \mu} = \beta_0 \p_0 B_{r \mu} \ .
\end{gather} 

We now introduce ghost partners $c_{r,a}$ 
for $\vp_{r,a}$ respectively, and require the action (in the absence of background fields) to be invariant under transformation 
\be \label{brst00}
\de \vp_r = \ep c_r, \qquad \de c_a = \ep \vp_a  \ .
\ee
In the presence of external sources it is convenient to introduce ghost partners $ \eta_{r \mu},  \eta_{a \mu}$
for  $A_{r \mu} , A_{a \mu}$ respectively and the action should be now be invariant under the combinations of~\eqref{brst00} and\footnote{See Appendix~\ref{app:a}  for motivation for introducing ghost partners and BRST transformations for external sources.}
\be \label{brst1}
\de A_{r \mu} = \ep \eta_{r \mu}, \qquad \de \eta_{a \mu} = \ep A_{a \mu}  \ .
\ee
Introducing 
\be 
H_{r \mu} = \eta_{r \mu} + \p_\mu c_r , \qquad H_{a \mu} = \eta_{a \mu} + \p_\mu c_a  \ .
\ee
then~\eqref{brst00}--\eqref{brst1}  can be written in a unified way as 
\be \label{1br}
\de B_{r \mu} = \ep H_{r \mu}, \qquad \de H_{a \mu} = \ep B_{a \mu}  \ .
\ee

Extending the dynamical KMS transformation~\eqref{ckmsn} to ghost fields proceeds in an identical manner as the example of Sec.~\ref{sec:exm} and we find 
\be
\tilde{H}_{a\mu}(x)=H_{r\mu}(-x) , \qquad \tilde{H}_{r\mu}(x)= - H_{a\mu}(-x) \ , 
\ee
or in terms of source and dynamical fields separately
\be
\tilde{\eta}_{r\mu}(x)=- \eta_{a\mu}(-x), \quad \tilde{\eta}_{a\mu}(x)=\eta_{r\mu}(-x), \quad \tilde{c}_{r\mu}(x)=c_{a\mu}(-x), \quad \tilde{c}_{a\mu}(x)=- c_{r\mu}(-x) \ .
\ee
Similarly one finds $\bar Q$ transformation is given by
\be
\bar{\de}B_{r\mu}=-\bar{\ep}H_{a\mu}, \quad \bar{\de}B_{a\mu}=\bar{\ep}i\beta_0 \p_0 H_{a\mu}, \quad \bar{\de}H_{r\mu}=\bar{\ep}(B_{a\mu}+i\beta_0 \p_0 B_{r\mu}) , \quad \bar \de H_{a \mu} = 0  \ .
\ee

Note that the above transformations for $B_{r \mu}, B_{a \mu}, H_{r \mu}, H_{a \mu}$ are identical to those of Sec.~\ref{sec:exm} for $\chi_{r \rmi},  \chi_{a \rmi}, c_{r \rmi}, c_{a \rmi}$. Thus all the results there can be directly carried over with simple change of notations. For example, the superfield now has the form 
\be
\Sig_\mu(x,\th,\bar{\th})=B_{r\mu}+\th H_{r\mu}+H_{a\mu}\bar{\th}+\t \bar{\th} B_{a\mu} \ .
\ee
The proof of the supersymmetrizability theorem also carries over as in the proof the nature of species indice $\rmi, \rmj$ 
did not play any role. Here they are replaced by $\mu, \nu$ of vector indices.

\section{BRST and emergent supersymmetry for fluctuating hydrodynamics} \label{sec:fluid}

As an example with both conserved quantities and dynamical temperature, in this section we consider the full fluctuating hydrodynamics for a relativistic charged fluid in the classical limit~\cite{CGL,CGL1}, which is the low energy effective theory for slow modes associated with stress tensor and a conserved $U(1)$ current.  

Here the story is much more complicated than those of the previous two examples. Remarkably, we will see in the end an almost identical structure to that of previous examples emerges when the theory is expressed 
in terms of an appropriate set of variables. There is also an important difference. In previous two examples with a fixed temperature we saw that the background temperature plays an important role 
in the supersymmetric algebra~\eqref{susya}. Now with a dynamical temperature we will see that supersymmetry becomes 
local.  

Supplementing the bosonic story of~\cite{CGL,CGL1} with the ghost sector, the discussion here completes the formulation of fluctuating hydrodynamics in the classical regime.

\subsection{Bosonic sector}  \label{sec:hydro}

The dynamical variables are given by  $\chi_{r} = (X^\mu (\sig), \vp_r (\sig), \beta (\sig))$ and $\chi_{a}= (X^\mu_a (\sig), \vp_a (\sig))$ with $\mu$ the spacetime index. Here $\sig^\al = (\sig^0, \sig^i)$ with $i=1,\cdots d-1$ are coordinates of  a ``fluid spacetime" labelling fluid elements and their internal clocks. $X^\mu (\sig^\al)$ gives ``physical'' spacetime coordinates $X^\mu$  of a fluid element labelled by $\sig^\a$ as in the standard Lagrange description of  fluid flows,  with $X^\mu_a$ describing the corresponding noises. 
As in the diffusion example of Sec.~\ref{sec:diff} $\vp_{r,a}$ are the charge diffusion mode associated with $U(1)$ and the corresponding noise. $\beta (\sig)$ is the local inverse temperature. We can write 
\be \label{tay0}
\beta (\sig) = {1 \ov T(\sig)} =  \beta_0 e^{\tau (\sig)} 
\ee
where $T_0 = {1 \ov \beta_0}$ is the temperature at infinities where we take all external sources and dynamical fields to vanish.
 $\beta_0$ is  the parameter appearing in the KMS condition~\eqref{1newfdt1}. 
 
The external sources are $g_{\mu \nu} (x), A_\mu (x)$ and 
$g_{a \mu \nu} (x), A_{a \mu} (x)$, with $g_{\mu \nu} (x)$ the spacetime metric.  They are defined in physical spacetime. Here $x \equiv x^\mu$ denotes physical spacetime coordinates and should be distinguished from dynamical variables $X^\mu (\sig)$.

The theory can be formulated either in fluid spacetime as the above variables indicate or in physical spacetime by inverting $X^\mu (\sig^\a)$. Below we will work in the fluid spacetime as it is more convenient for introducing ghost partners and writing down a supersymmetric action. 

Quantities can be pulled-back or pushed-forward between the fluid and physical spacetimes through $X^\mu (\sig^\al)$.  
For example, from $\vp_a (\sig^\a)$ we can obtain $\vp_a' (x) = \vp_a (\sig (x))$ where $\sig^\al (x^\mu)$ is the inverse function of  $X^\mu (\s)$.  Unless otherwise specified  below for notational simplicity we will always use the same notation for a quantity 
and its push-forward (or pull-back), i.e.  write $\vp_a' (x) $ simply as $\vp_a (x)$ and distinguish $\vp_a (x)$ from $\vp_a (\s)$ either by its argument or from context.

In the classical limit the action depends on the dynamical variables and external sources only through certain combinations, more explicitly (for more details see~\cite{CGL1})
\be
 I_{\rm hydro}  = I_{\rm hydro} [h_{\a \b}, B_\a, \beta^\a; h_{a \a \b}, B_{a\a}] 
 \ee
where 
\bega\label{pp1}
h_{\a \b} (\sig) \equiv  \p_\a X^\mu\p_\b X^\nu g_{\mu \nu} (X) , \qquad h_{a \a\b} = \p_\a X^\mu\p_\b X^\nu (g_{a\mu \nu} + \sL_{X_a} g_{\mu \nu} ), \\
\label{pp11}
B_\a  \equiv  \p_\a X^\mu A_\mu (X) + \p_\a \vp (\sig), \qquad B_{a \a}  = \p_\a X^\mu (  A_{a\mu}(X )+ \sL_{X_a} A_\mu )
+ \p_\al \vp_a 
%  C_{a \mu}  (X) \\
%G_{a\mu\nu} (X) 
% \equiv g_{a\mu \nu} + \sL_{X_a} g_{\mu \nu} 
%, \qquad
%C_{a\mu} \equiv  A_{a\mu}(X )+ \p_\mu \vp_a (X) + \sL_{X_a} A_\mu \ ,
%\label{pp2}
\end{gather} 
with %$\vp_a (X) \equiv \vp_a (\sig (X))$.  
$\sL_{X_a}$ denotes the Lie derivative along the vector $X_a^\mu (x) \equiv X_a^\mu (\sig (x))$. 
In addition to~\eqref{fer1}--\eqref{key1}, the action should also be invariant under separate spatial and time diffeomoprhisms 
\be  \label{diff} 
\sig^i \to \sig'^i (\sig^i), \qquad \sig^0 \to \sig'^0 (\sig^0, \sig^i)
\ee
as well as~\eqref{csh}. 
The local velocity and chemical potential are defined as 
\be 
u^\mu  = {1 \ov \sqrt{-h_{00}}} \p_0 X^\mu , \qquad \mu = {1 \ov \sqrt{-h_{00}}}  B_0  \ .
\ee
We also introduce a local temperature vector in fluid spacetime
\begin{equation} \label{defb}
 \beta^{\a}\equiv\f{\beta (\sig)}{\sqrt{-h_{00}}}\left(\f{\del}{\del\s^{0}}\right)^{\a} = \f{\beta_0 e^\tau}{\sqrt{-h_{00}}}\left(\f{\del}{\del\s^{0}}\right)^{\a}
 ,\qquad
h_{\al \b} \b^{\a}\b^{\b}=-\b^{2} \ 
\end{equation}
and it push-forward in physical spacetime 
\be 
\beta^\mu (x)= \p_\al X^\mu \beta^\a = \beta (x) u^\mu (x)   , \quad  %u^\mu = {1 \ov \sqrt{-h_{00}}} \p_0 X^\mu , \quad
g_{\mu \nu} \beta^\mu \beta^\nu = - \beta^2  \ .
\ee

The dynamical KMS transformation on bosonic variables can be written as~\cite{CGL1} 
\bega \label{x1}
\tilde X^\mu (\sig) = - X^\mu (-\sig), \qquad \tilde X^\mu_a (\sig) = - X_a^0 (-\sig)  
-i  \beta^\mu (-\sig) + i \beta_0^\mu 
 \\
\tilde \vp  (\sig) = - \vp (-\sig), \quad \tilde \vp_a (\sig) = - \vp_a (-\sig)- 
 i \beta^\a \p_\a  \vp (-\sig) , \quad \tilde \b (\sig) = \b (-\sig), \quad 
 \label{x2}
\end{gather}  
with $\beta_0^\mu = \beta_0 \de_0^\mu$ and 
\bega 
\tilde g_{\mu \nu} (x) = g_{\mu \nu} (-x), \quad 
\tilde g_{a \mu \nu} (x) =  g_{a \mu \nu} (-x) + i \sL_{\beta_0} g_{\mu \nu} (-x), \\
\tilde A_{\mu } (x) = A_{\mu } (-x)  , \quad  \tilde A_{a\mu } (x) = A_{a\mu } (-x) +  i \sL_{\beta_0} A_{\mu } (-x) \ .
\end{gather}
The quantities in~\eqref{pp1}--\eqref{pp11} then transform as
\bega \label{pi1}
 \tilde h_{\a\b} (-\sig) =  h_{\a\b} (\sig) , \quad  \tilde \b^\a (-\s) = \b^\a (\sig), \quad
\tilde h_{a \a\b} (-\sig) = h_{a\a\b} (\sig) + i  \sL_{\beta} h_{\a\b} (\sig) \\
 \tilde B_{\a} (-\sig) = B_{\a} (\sig) , \qquad \tilde B_{a\a} (-\sig) = B_{a\a} ( \sig) + i  \sL_{\beta} B_{\a} (\sig) 
 \label{pi2}
\end{gather}
where $\sL_{\beta}$ denotes the Lie derivative along the vector $\beta^\a$.

For our discussion of BRST symmetry below, we will need to use vielbeins\footnote{
Below expressions come from $\hbar \to 0$ limit of  $g_{1 \mu \nu} = \eta_{AB} e_{1 \mu}^A e_{1 \nu}^B$ and $g_{2 \mu \nu} = \eta_{AB} e_{2 \mu}^A e_{2 \nu}^B$ with $g_{1 , 2 \mu \nu}  = g_{\mu \nu} \pm {\hbar \ov 2} g_{a \mu \nu}$ and $e_{1,2 \mu}^A = e_{\mu}^A \pm {\hbar \ov 2} e_{a \mu}^A$. %$g_2$ and $e_2$ are expanded with the signs of the corresponding $a$-field terms flipped.
} for various quantities in~\eqref{pp1} 
\bega
 g_{\mu\nu}  =e_{\mu}^{A}e_{\nu}^{B}\eta_{AB}, \quad g_{a \mu\nu}  =\le(e_{a\mu}^{A} e_{\nu}^{B} + e_{\mu}^{A} e_{a\nu}^{B} \ri) \eta_{AB}, \quad
 h_{\a\b}  =f_{\a}^{A}f_{\b}^{B}\eta_{AB},  \\
   f_{\a}^{A}  =\del_{\a}X^{\mu}e_{\mu}^{A} , \quad h_{a\a\b}  =\le( f_{a\a}^{A} f_{\b}^B + f_{\a}^{A} f_{a \b}^B\ri)  \eta_{AB}, \quad
  f_{a \a}^{A}  =\del_{\a}X^{\mu}\left(e_{a\mu}^{A}+\mL_{X_a}e_{\mu}^{A}\right) \ .
  \label{pp12}
\end{gather}
Under dynamical KMS transformation we have 
\bega
\tilde e_\mu^A (-x) = \tilde e_\mu^A (- x) , \qquad \tilde e_{a \mu}^A (-x)  = e_{a \mu}^A (x) + i \sL_{\beta_0} e_\mu^A (x) \\
\tilde f_\al^A (-\sig) = f_\al^A (\sig), \qquad  \tilde{f}_{a \a}^{A}(- \s)=f_{a \a}^{A}(\s)+i\mL_{\b}f_{\a}^{A}(\sig) \ .
\label{eq:28}
\end{gather} 
The action can thus also be considered as
\be \label{acd}
I_{\rm hydro} [f_\a^A, f_{a \a}^A, B_\a, B_{a \a}, \beta^\a]  \ . 
\ee
Note that in writing down explicit terms we will often need to use  the inverse $f^\a_A$ of $f_\a^A$, whose transformation can be worked out from the above. 
%which can be expressed as an algebraic functional of $f^A_\a$. 

\subsection{BRST transformations and supersymmetry} \label{sec:bgg}

We now introduce ghost partners. For dynamical variabless,  $\ga^\mu, \ga_a^\mu$ for $X^\mu, X_a^\mu$, and $c_{r,a}$ for $\vp_{r,a}$. For external fields,  $m_\mu^A, m_{a \mu}^A$ for $e_\mu^A , e_{a \mu}^A$, and $\eta_{ \mu}, \eta_{a \mu}$ for $A_{\mu}, A_{a \mu}$. The BRST transformations again follow~\eqref{obrst} and are given by 
\bega \label{ba1}
\d X^{\mu}=\e\g^{\mu} , \qquad  \d\g_{a}^{\mu}=\e X_{a}^{\mu},  \qquad \d e_{\mu}^{A} (x) =\e m_{\mu}^{A}  (x) , \qquad
\d m_{a\mu}^{A} (x) =\e e_{a\mu}^{A} (x) \\
\d\vp_{r}=\e c_{r}, \qquad \d c_{a}=\e\vp_{a}, \qquad \d A_{\mu} (x) =\e\eta_{\mu} (x), \qquad 
\d\eta_{a\mu} (x) =\e A_{a\mu} (x) \ .
\label{ba2}
\end{gather} 
We should stress that the above transformations for $e_{\mu}^{A}$, $A_{\mu}$, $\eta_{a\mu}$ and $m_{a\mu}^{A}$ are defined in physical spacetime. When pulled back to fluid spacetime we then have, for example,  
\bega 
\de e_\mu^A (X(\sig)) = \de X^\nu \p_\nu e_\mu^A (X (\sig)) + \ep m_\mu^A (X (\sig)) = 
\ep (m_\mu^A (X (\sig)) + \ga^\nu (\sig) \p_\nu e_\mu^A (X (\sig))) \\
\Rightarrow \qquad \de f_a^A = \ep \p_\a X^\mu (m_\mu^A + \sL_\ga e_\mu^A ) \ .
\end{gather}

We do {\it not} introduce any ghost partner for $\b (\sig)$ and require the BRST transformation of $\b$ be such
that $\d\b^{\a}=0$:
\begin{align}
\d\b^{\a}  =\d_{0}^{\a}\b \left(\d\tau(-h_{00})^{-1/2}+\d(-h_{00})^{-1/2}\right)=0 %\nonumber \\
\quad \implies \quad \d \log \b   = \de \log \sqrt{-h_{00}} \ . % -(-h_{00})^{1/2}\d(-h_{00})^{-1/2}
\label{eq:13}
\end{align}
Note that the transformation of $\beta$ is complicated as $h_{00}$ consists of that of $X^{\mu}$
and $e_{\mu}^{A}$.  From now on we can just simply regard $\b^{\a}$ 
as a BRST invariant field. 
As a result we have 
\be 
\de \b^\mu (\sig) =\ep  \beta^\al \p_\a \ga^\mu (\s), \qquad \de \b^\mu (x) = \ep \sL_{\b} \ga^\mu (x) \ .
\ee

Since in the action~\eqref{acd} only $f_\a^A, f_{a\a}^A, B_\a, B_{a \a}$ can appear, %but these quantities themselves do not all transform nicely under~\eqref{ba1}--\eqref{ba2}. 
let us now construct objects which contain them and at the same time have good transformation properties under~\eqref{ba1}--\eqref{ba2}. This process is facilitated by considering BRST superfields defined as follows:
\bega \label{o1}
\mX^{\mu}=X^{\mu}+\t\g^{\mu}, \quad 
 \mX_a^{\mu}=\g_{a}^{\mu}+\t X_{a}^{\mu} , \quad \sE_{\mu}^{A}  =e_{\mu}^{A}+\t m_{\mu}^{A}, \quad \sE_{a \mu}^{A} =m_{a \mu}^{A}+\t e_{a \mu}^{A},  \\
 \Phi=\vp_{r}+\t c_{r}, \quad \Phi_a=c_{\a}+\t\vp_{\a}, \quad 
\mA_{\mu}=A_{\mu}+\t\eta_{r\mu},  \quad 
\mA_{a \mu}=\eta_{\a\mu}+\t A_{\a\mu} \ .
\label{o3}
\end{gather}
Motivated from~\eqref{pp11} and~\eqref{pp12} now consider 
\bega \label{oo1}
 \sF_{\a}^{A}  = \del_{\a}\mX^{\mu} \sE_{\mu}^{A}(\mX) \equiv  f_{\a}^{A}+ \th M_{\a}^{A} , \qquad 
 \mB_{\a} = \del_{\a}\mX^{\mu}\mA_{\mu}(\mX)+\del_{\a}\Phi  \equiv B_{\a}+\t H_{\a},  \\
 \sF_{a \a}^{A}  = \del_{\a}\mX^{\mu} \le(\sE_{a \mu}^{A}(\mX)+ \sL_{\mX_a} \sE_\mu^A (\mX) \ri) \equiv M_{a \a}^A + \th F_{a \al}^A
  \\
  \mB_{a\a }  = \del_{\a}\mX^{\mu}\le( \sA_{a \mu}(\mX) + \sL_{\mX_a} \sA_\mu \ri) +\del_{\a} \Phi_a 
  \equiv H_{a \a} + \th \mathfrak{B}_{a \al}
  \label{oo3}
  \end{gather}
where 
\be 
\del_{\a}\mX^{\mu} \sL_{\mX_a} \sA_\mu (\mX)  \equiv  \del_{\a}\mX^{\mu} \mX^\nu_a \p_\nu \sA_\mu (\mX) + 
\p_\a \mX_a^\mu \sA_\mu (\mX) 
\ee
and similarly with $ \del_{\a}\mX^{\mu} \sL_{\mX_a} \sE_\mu^A (\mX)$.  Various quantities in~\eqref{oo1}--\eqref{oo3} are given by 
\bega
M_{\a}^{A} = \del_{\a}X^{\mu}(m_{\mu}^{A}+\mL_{\g}e_{\mu}^{A}), \qquad 
M_{a \a}^A = \del_{\a}X^{\mu}(m_{a\mu}^{A}+\mL_{\g_{a}}e_{\mu}^{A}) \\
F_{a \al}^A =f_{a \a}^{A}+\del_{\a}X^{\mu}(\mL_{\g}m_{a\mu}^{A}-\mL_{\g_{a}}m_{\mu}^{A}+\mL_{\g \g_{a}} e_{\mu}^{A}) \\
H_\al = \del_{\a}X^{\mu}(\eta_{r\mu}+\mL_{\g}A_{\mu})+\del_{\a}c_{r}, \qquad H_{a \al} = 
\del_{\a}X^{\mu}(\eta_{a\mu}+\mL_{\g_{a}}A_{\mu})+\del_{\a}c_{a} \\
\mathfrak{B}_{a \al} = B_{a \a}+\del_{\a} X^{\mu}(\mL_{\g} \eta_{a\mu}-\mL_{\g_{a}} \eta_{r\mu}+\mL_{\g \g_{a}}A_{\mu}) \ .
\end{gather} 
In the above expressions $\sL_\ga$, $\sL_{\ga_a}$ are defined as usual, while 
\bea
\del_{a}X^{\mu}\mL_{\g\g_{a}}V_{\mu} &\equiv & \del_{a}X^{\mu}\g^{\nu}\g_{a}^{\rho}\del_{\nu}\del_{\rho}V_{\mu}+\g^{\mu}\del_{a}\g_{a}^{\nu}\del_{\mu}V_{\nu}+\del_{a}\g^{\mu}\g_{a}^{\nu}\del_{\nu}V_{\mu}  \cr
&= &\frac{1}{2}\del_{a}X^{\mu} \le([\mL_\g, \mL_{\g_a}] - \mL_{[\ga, \ga_a]} \ri) V_\mu 
\eea
which is covariant explicitly. 
We thus find that the BRST extensions of $f_{a\al}^A, B_{a \al}$ are  respectively $F_{a \al}^A, \mathfrak{B}_{a \al} $ and 
\be \label{newk}
\de f_\a^A = \ep M_\a^A , \quad \de M_{a \a}^A = \ep F_{a \a}^A, \quad  \de B_\a = \ep H_\a, \quad \de H_{a \a} = \ep \mathfrak{B}_{a \al} \ .
\ee

%whose partners are $M_{a \al}^A, H_{a \al}$ respectively.  The partners of $f_\a^A, B_\a$ are  $M_\al^A$ and $H_\a$.  The action should be constructed from these quantities plus $\b^\a$. 

%\subsection{Dynamical KMS transformation on ghosts and supersymmetry} 

Let us now consider the dynamical KMS transformation on ghost variables. 
As in previous examples, by examining the quadratic action we propose that 
\bega \label{ca1}
\tilde{\g}_{a}^{\mu}(\s)= - \g_{r}^{\mu}(-\s), \quad  \tilde{\g}^{\mu}(\s)= \g_{a}^{\mu}(-\s), \quad
\tilde{c}_{a}(\s) = -  c_{r}(-\s), \quad 
\tilde{c}_{r}(\s)= c_{a}(-\s) , \\
\tilde{m}_{a\mu}^{A}(x) = m_{\mu}^{A}(-x ), \;\; \tilde{m}_{\mu}^{A}(x) = - m_{a\mu}^{A}(-x ), \;\;
\tilde{\eta}_{a\mu} (x) = \eta_{\mu} (-x ), \;\; \tilde{\eta}_{\mu}(x) = - \eta_{a\mu}(-x ) \ .
\label{ca2}
\end{gather} 
Again the transformations of sources are given in physical spacetime and when pulled back to the fluid spacetime 
the arguments should also transform, e.g. 
\be
m_{a\mu}^{A}(X(\s)) \quad  \ra \quad \tilde{m}_{a\mu}^{A}(\tilde{X}(\s))= m_{\mu}^{A}(-\tilde{X}(\s))=m_{\mu}^{A}(X(-\s))
\ee
where we have used~\eqref{x1}. Applying~\eqref{x1}--\eqref{x2} and~\eqref{ca1}--\eqref{ca2} to~\eqref{oo1}--\eqref{oo3}  we find that 
\bega\label{newk1}
\tilde f_\a^A (-\sig) = f_\a^A (\s), \qquad \tilde F_{a \al}^A (-\s) = F_{a \al}^A (\s) + i \sL_{\beta} f_\a^A (\s) \\
 \tilde M_\a^A (-\sig) = - M_{a \a}^A (\sig), \qquad \tilde M_{a\a}^A (-\sig) = M_{\a}^A (\sig), \\
 \tilde B_\a (-\sig) = B_\a (\s), \qquad \tilde{\mathfrak{B}}_{a \al} (-\s) = \mathfrak{B}_{a \al} (\s) + i \sL_{\beta} B_\a (\s) \\
 \tilde H_\a (-\sig) = - H_{a \a} (\sig), \qquad \tilde H_{a\a}  (-\sig) = H_{\a} (\sig) \ . 
 \label{newk4}
\end{gather} 

From~\eqref{newk} and~\eqref{newk1}--\eqref{newk4} we see that the multiplets $(f_\a^A, F_{a\al}^A, M_\a^A, M_{a \a}^A)$ and $(B_\a, \mathfrak{B}_{a \a}, H_\a, H_{a \a})$ have identical structure in terms of BRST and dynamical KMS transformations as $(\chi_{r \rmi}, \chi_{a \rmi}, c_{r \rmi}, c_{a \rmi})$ 
of Sec.~\ref{sec:exm} with the replacement of $i \beta_0 \p_0 $ by $i \sL_{\beta}$. Thus all the subsequent discussion there regarding supersymmetry can be carried over immediately. In particular,
the superalgebra~\eqref{susya} becomes 
\be\label{susyb}
\{Q, \bar Q\} =  i \sL_\b \  
\ee
and the action can be constructed using the following two superfields 
\be
\Lam_\a^A (\s, \th, \bar \th) = f_{\a}^A +\th M_{\a}^A+M_{a\a}^A \bar{\th}+\th \bar{\th} F_{a\a}^A  , \quad \Sig_\a(\s,\th,\bar{\th})=B_{\a}+\th H_{\a}+H_{a\a}\bar{\th}+\th \bar{\th} \mathfrak{B}_{a\a} \ 
\ee
and their (super)-derivatives.  

In contrast to the examples of last two sections here the right hand side of the supersymmetric algebra~\eqref{susyb}
depends on dynamical fields.  In particular, $\b^{\a}$ contains some complicated dependence on $f_\a^A$ (recall~\eqref{defb}).  
Thus a supersymmetric transformation no longer preserves the total power of  fields, so 
the proof of supersymmetrization theorem of Sec.~\ref{sec:the} and Appendix~\ref{app:proof} 
cannot be immediately applied. 
This potential problem can be avoided as follows. By using the freedom of time reparameterization~\eqref{diff} we can 
set $\beta^\al = \beta_0 \de^\al_0$ by choosing 
\be \label{ikm}
\sqrt{-h_{00}} = e^\tau , \quad i.e. \quad f_0^A f_0^B \eta_{AB} = - e^{2 \tau} \ .
\ee
Then the supersymmetry 
becomes global, i.e. 
\be 
\{Q, \bar Q\} = i \beta_0 \p_0 
\ee
and  the proof can be applied at least when expanding the action around an equilibrium configuration.
Note that  in this gauge local temperature is expressed through $f_0^A$ via~\eqref{ikm}.

\section{Conclusions and discussions} \label{sec:conc}

We first summarize the main results of the paper. Formulating  a consistent non-equilibrium effective field theory for a system in local equilibrium 
requires imposing BRST and dynamical KMS symmetries. 
We showed that BRST and dynamical KMS symmetries always lead to 
an emergent supersymmetry. Conversely, supersymmetry provides a convenient way to impose BRST symmetry and the special dynamical KMS condition. Starting from a supersymmetric action one can then construct a BRST and dynamical KMS  invariant theory through a simple procedure~\eqref{susm}. We have discussed a few explicit examples in detail, in particular completing the formulation of fluctuating hydrodynamics of~\cite{CGL,CGL1} by understanding how to introduce ghosts and implement various symmetries in the ghost sector. %It is straightforward to generalize our discussion to more complicated systems involving slow variables associated with both non-served and conserved quantities. 

Our results have important implications for studying non-equilibrium questions. Ghosts run in loops and 
thus their dynamics % full dynamical KMS invariance 
plays an important role in %constraining the dynamics of a system. when considering loop corrections. In other words, 
understanding physical effects of statistical fluctuations on a physical process or physics observables. 
For example, it would be interesting to re-examine various dynamical critical phenomena in this light. With a full action for
fluctuating hydrodynamics, one could explore systematically many questions related to effects of fluctuations on transport coefficients, hydrodynamic instabilities, and so on, in particular in far-from-equilibrium situations.  At a technical level supersymmetry may also help find  nontrivial fixed points for such non-equilibrium effective field theories. 

%See e.g.~\cite{Kovtun:2011np,Kovtun:2012rj,Gripaios:2014yha,Endlich:2010hf} for some recent discussions of the effects of fluctuations.

There are still a number of conceptual and technical challenges to overcome in order to generalize the current discussion to the quantum regime. 
At conceptual level, quantum fluctuations operate at the scale of ${\hbar \ov T}$,\footnote{Here we are having in mind a strongly coupled system which is of our main interests.} which is essentially the cutoff scale for an EFT,  and thus makes  the theory intrinsically nonlocal. This can already be seen from~\eqref{ndym} which involves a translation of order ${\hbar \ov T}$, therefore  requiring that the theory should be able to resolve such a scale. Technically the transformation involves an infinite number of derivatives and thus invalidates derivative expansion. While one could take into quantum effects perturbatively by developing an expansion in $\hbar$, it is an interesting question whether it is possible to capture full quantum effects at low energies by relaxing locality. Indeed at quadratic order  around thermal equilibrium it was found in~\cite{CGL} that one could write down an ``effective field theory'' which includes an infinite number of derivatives. In particular, in this theory  BRST and dynamical KMS symmetries lead to a quantum deformed supersymmetric algebra
  \be \label{qun}
\{Q, \bar Q\} =  {2 \ov \hbar} \tanh {i \hbar \beta_0  \p_t \ov 2} \ .
\ee
There are immediate technical difficulties in generalizing such an algebra to nonlinear level, as acting on a finite product of
local fields the left hand side is a derivation while the right hand side is not. Finally there is a potentail ambiguity in the dynamical KMS transformations for hydrodynamical variables at quantum level~\cite{CGL1}. We hope to return to these issues in the future.

\vspace{0.2in}   \centerline{\bf{Acknowledgements}} \vspace{0.2in}
We thank  P.~Glorioso, K.~Jensen, and A.~Yarom for discussions. Work supported in part by funds provided by the U.S. Department of Energy
(D.O.E.) under cooperative research agreement DE-FG0205ER41360.

\appendix

\section{BRST symmetry and external sources} \label{app:a}

In this appendix we review the argument that BRST symmetry ensures~\eqref{nor}. 

Consider an action $I_b [\chi_r, \phi_r; \chi_a, \phi_a]$ where $\chi_{r,a}$ denote collectively the dynamical variables 
and $\phi_{r,a}$ external sources. Now let us introduce BRST partners $c_{r,a}$ for dynamical fields and 
require the full action in the presence of $\phi_r$ (with $\phi_a =0$) to be invariant under 
\be \label{qqi}
\de \chi_r = \ep c_r, \qquad \de c_a = \ep \chi_a \ .
\ee
We can write this as\footnote{In this case we can always write a $Q_d$-closed quantity as a $Q_d$-exact.}
\be \label{qi}
Q_d I = 0 , \qquad  I = Q_d \sF , \qquad Q_d = c_r {\de \ov \de \chi_r} +  \chi_a {\de \ov \de c_a}  
\ee
for some $\sF$ where $I$ denotes the total action including ghosts. From~\eqref{qi} under variation of $\phi_r$ we then have%\footnote{Note that the first equation of~\eqref{qi} only implies that $Q_d {\de I \ov \de \phi_r} =0$ which is weaker than~\eqref{qi1}.} 
\be \label{qi1}
{\de I \ov \de \phi_r}=  Q_d V 
\ee
for some operator $V$ and  under a variation of $\phi_r$ the full generating functional is 
\be\label{argd}
e^W {\de W }   =  i \int  D \chi_r D \chi_a D c_r D c_a \, \le(Q_d   V \ri)\, e^{i I}
 =  i \int D \chi_r D \chi_a D c_r D c_a \, Q_d   \le( V e^{i I}  \ri)
 =0,
\ee
where in the second equality we have used that $I$ is BRST invariant and in
the third equality we have used that $Q_d$ as defined in~\eqref{qi} is  a total derivative under the path integration.
We have thus shown that $W$ is independent of $\phi_r$ and can then be set to zero by choosing a normalization constant. 

To ensure the action is BRST invariant for any $\phi_r$, it is convenient to introduce also BRST partners $\eta_{r,a}$ for external sources $\phi_{r,a}$ and require the full action (with all external fields turned on) to be invariant under simultaneous transformation of~\eqref{qqi} and the corresponding transformations on sources  
\be 
\de \phi_r = \ep \eta_r, \qquad \de \eta_a = \ep \phi_a  \ .
\ee
We thus have 
\be 
Q I = 0, \qquad Q = c_r {\de \ov \de \chi_r} +  \chi_a {\de \ov \de c_a}  +  \eta_r {\de \ov \de \phi_r}  + \phi_a {\de \ov \de \eta_a}
= Q_d + Q_s
\ee  
where $Q_s$ denotes the source part of the BRST operator. Now setting $\eta_{r,a}$ and $\phi_a$ to zero we then obtain an action which is  automatically invariant under $Q_d$ for all $\phi_r$. 

For the examples discussed in the main text, equation~\eqref{argd} clearly applies to model A and nonlinear diffusion with conserved quantities. For fluctuating hydrodynamics, 
there is an interesting subtlety. To guarantee physical spacetime diffeomorphisms $X \ra X'(X)$, the path integral measure for $X$'s should come in the form $DX\sqrt{G(X)}$, in which case there is then a nontrivial variation $\d \sqrt{G(X)}/\d X$ inside the path integral and~\eqref{argd} appears to break down. But for the effective action to be physical spacetime diffeomorphism invariant,  $X^\mu$'s fermionic partner $\g^\mu$  introduced in Sec.~\ref{sec:bgg} should also transform as a physical spacetime vector, i.e. $\g^\mu(X)\equiv\g^\mu(\s(X))\ra\g'^{\nu}(X')=\del_\mu X'^{\nu}(X) \g^\mu(X)$. 
Since the Jacobian for Grassmanian field $\g^\mu$ is the inverse of that of $X^\mu$, the measure $DXD\g$ is invariant under physical spacetime diffeomorphisms. The same cancellation of Jacobian exists between $X_a^\mu$ and its partner $\g_a^\mu$. Thus there is no need for including $\sqrt{G(X)}$ in the integration measure and \eqref{argd} applies.

%\textcolor{red}{PG: I have a subtle confusion. In the fluid case, to guarantee diffeomorphism in physical space, it seems that the path integral measure should be $DX_1 DX_2 \sqrt{\det(g_1(X_1))}\sqrt{\det(g_2(X_2))}$ besides the action to be diff-invariant. Since the determinant of metric involves dynamical fields, can we still conclude A4?}

\section{Proof of special KMS condition implying supersymmetrization \label{app:proof}}

In this Appendix we prove that the most general solution to equation~\eqref{skm3} has the form~\eqref{dfe}. 
The general proof is a bit involved and notation heavy. We will start by considering some simpler cases, which already captures the 
essence of the proof.

\subsection{Simplest case}

Let us first consider a simplest case for which there is only one species and all fields only depend on time (i.e. a quantum mechanical example). Equation~\eqref{skm3} can now be written explicitly as 
\be \label{uye}
\om_1 f(\om_1 ; \om_2 , \cdots ,\om_n) \chi_r (\om_1) \cdots \chi_r (\om_n) = 0  
\ee
where $f(\w_{1};\w_{2},\cdots,\w_{n})$
is a polynomial symmetric in last $n-1$ variables.  The above equation can be written more explicitly as 
\be \label{eq:16}
\om_1 f(\om_1 ; \om_2 , \cdots ,\om_n) + \om_2 f(\om_2 ; \om_1 , \cdots ,\om_n) + \cdots
+ \om_n f (\om_n; \om_2, \cdots , \om_{n-1},\om_1) =0 \ . 
\ee
Expanding $f$ in polynomials of $\om_i$, equation~(\ref{eq:16}) is valid degree by degree. Thus without loss of generality we can take  $f$ to have  degree $h$ and write it as
\begin{equation}
f=\w_{1}^{h}f_{0}+\w_{1}^{h-1}f_{1}+\cdots+\w_{1}f_{h-1}+f_{h}\label{eq:17}
\end{equation}
where $f_{k}$ is a symmetric homogeneous polynomial of $\w_{2},\cdots,\w_{n}$
of degree $k$.

The basic idea of the proof is to use symmetric polynomials to write $f$ in a suitable form which then enables us to solve~\eqref{eq:16} explicitly. 

Now introducing the power sum basis for symmetric homogeneous polynomials of  $\w_{1},\cdots,\w_{n}$ %as $q_{k}$ symmetric 
\begin{equation} \label{qk}
q_{k}=\sum_{s=1}^{n}\w_{s}^{k},\qquad k=0,\cdots,n
\end{equation}
with some manipulations of symmetric polynomials we can further write $f$ as 
\begin{equation}
f=\w_{1}^{n-1}g_{h-n+1}(q_{k})+\w_{1}^{n-2}g_{h-n+2}(q_{k})+\cdots+\w_{1}g_{h-1}(q_{i})+g_{h}(q_{k})
\label{eq:23-1}
\end{equation}
where $g_m$ is a symmetric homogeneous polynomial of $\w_{1},\cdots,\w_{n}$
of degree $m$ and can be expanded in basis $\{q_k\}$ as already indicated in~\eqref{eq:23-1}. 
We stress that in contrast to~\eqref{eq:17}, the powers of $\om_1$ in~\eqref{eq:23-1} have been lowered to have maximal value $n-1$ and $g_m$ are symmetric in all frequencies. 
In order to focus on the main idea we leave the justification of~\eqref{eq:23-1} to Appendix~\ref{app:c4}. 

Expanding various functions $g_m$ in (\ref{eq:23-1}) in $q_i$ we can write $f$ as 
\begin{equation}
f=\sum_{P} \left(a^{(P)}_1  \w_{1}^{n_{1}-1}q_{n_{2}}\cdots q_{n_{k}}+a^{(P)}_2 q_{n_{1}}\w_{1}^{n_{2}-1}\cdots q_{n_{k}}+\cdots+a^{(P)}_k q_{n_{1}}\cdots q_{n_{k-1}}\w_{1}^{n_{k}-1}\right)
\label{eq:23}
\end{equation}
where $P = \{n_1 , \cdots, n_k\}$ denotes a (non-ordered) partition of $h+1$ (i.e. $\sum_{i=1}^k n_{i}=h+1$) 
and the sum is over all partitions (with all possible $k$ and $\{n_i\}$). Now consider a partition $P$ 
with at least one $n_j =1$, for which the corresponding terms in~\eqref{eq:23} are of two types:  
one is $q_{n_1}\cdots q_{n_{j-1}}q_{n_{j+1}}\cdots q_{n_{k}}$
and the other is proportional to $q_{1}$. The former is fully symmetric in all the $\om_s$'s and thus belongs to the first term 
in~\eqref{dfe}. The latter vanishes by momentum conservation. We thus find that 
\be \label{b4}
f = f^{(s)} + f^{(a)}, \qquad f^{(a)} =  \sum_{P'} \le(\cdots \ri)
\ee
where $f^{(s)}$ is fully symmetric in all $\om_s$'s and $P'$ denotes those partitions  with $n_{i}>1$ for all $i$. Substituting
(\ref{b4}) into~(\ref{eq:16}) we find 
\be 
\sum_{P'}  \le(a^{(P')}_1 + a^{(P')}_2 + \cdots + a^{(P')}_k \ri) 
q_{n_{1}}q_{n_{2}}\cdots q_{n_{k}}  = 0 \label{eq:24}
\ee
where the term containing $f^{(s)}$ vanishes by momentum conservation. Since $q_{n_{1}}q_{n_{2}}\cdots q_{n_{k}}  $ for different partitions are independent we thus have 
\be 
- \le(a^{(P')}_1 + a^{(P')}_2 + \cdots + a^{(P')}_{k-1} \ri) = a^{(P')}_k 
\ee
for all partitions $P'$. Substituting the above equation back into $f^{(a)}$ we find that 
\begin{equation}
f^{(a)} = \sum_{P'} \sum_{j=1}^{k-1}a^{(P')}_j q_{n_{1}}\cdots\cancel{q_{n_{j}}}\cdots q_{n_{k-1}}(\w_{1}^{n_{j}-1}q_{n_{k}}-q_{n_{j}}\w_{1}^{n_{k}-1})
\label{eq:26-1}
\end{equation}
where and below slash means omitting that term. Now putting~\eqref{eq:26-1} back into 
\be 
\sI^{(a)}_b = f^{(a)} \chi_a (\om_1) \chi_r (\om_2) \cdots \chi_r (\om_n)
\ee
and using permutations among $\chi_r$'s we can write 
\be
\sI^{(a)}_b  =  (n-1) \om_2 \sum_{P'} \sum_{j=1}^{k-1}a^{(P')}_j q_{n_{1}}\cdots\cancel{q_{n_{j}}}\cdots q_{n_{k-1}}(\w_{1}^{n_{j}-1}\w_2^{n_{k}-1}-\w_2^{n_{j}-1}\w_{1}^{n_{k}-1}) \chi_a \chi_r^{n-1}
\ee
which is the form of~\eqref{uei}.

\subsection{The next simpler case} \label{subsec:More-general-case:}

Now we still consider a single field species, but with full momentum 
$k^{\mu} = (\om, k^i), i =1, \cdots, d-1$ dependence in $d$ spacetime dimension, i.e. $f$ in~\eqref{uye} should be understood as 
$f(k_{1}^{\mu};k_{2}^{\mu},\cdots,k_{n}^{\mu})$ which is a polynomial of $nd$ variables $k_{s}^{\mu}$ and is symmetric
under permutations of $k_{s}^{\mu}$ for $s=2,\cdots,n$. Equation~\eqref{eq:16} now becomes 
\be \label{eq:160}
\om_1f(k_{1}^{\mu};k_{2}^{\mu},\cdots,k_{n}^{\mu}) + \om_2f(k_{2}^{\mu};k_{1}^{\mu},\cdots,k_{n}^{\mu})+ \cdots
+ \om_n  f(k_{n}^{\mu};k_{2}^{\mu},\cdots,k_{1}^{\mu}) =0 \ . 
\ee

The idea for solving~\eqref{eq:160} is exactly the same as in last subsection with the only difference being that 
we now need to use multi-symmetric polynomials that are natural generalizations of symmetric 
polynomials~\cite{dalbec1999multisymmetric,briand2004algebra}.

The generalization of a power $\om^m$ to multiple variables is $k^\al \equiv\w^{\alpha_{0}}(k^{1})^{\alpha_{1}}\cdots(k^{d-1})^{\a_{d-1}}$ where multiple power $\al$ is a $d$-dimensional vector in $\N^d$. 
We can choose a basis for $\al$-space as $e_{\mu}=(0,\cdots0,1,0,\cdots,0)$ for
$\mu=0,\cdots,d-1$ where $1$ is at $(\mu+1)$-th position. The multi-degree $h$ of a monomial
$\prod_{s=1}^n k_s^{\al_s}$ is defined as $h = \sum_{s=1}^n \al_s $ and is also a vector in  $\N^d$.
For example $\w_{1}^{2}\w_{2}\prod_{i=1}^{d-1} (k_{1}^{i}k_{2}^{i})$ has 
$\al_1 =(2,1,\cdots ,1) $, $\al_2 = (1,1,\cdots,1)$, $\al_s = (0, \cdots , 0)$ for $s=3, \cdots, n$ and $h = (3,2,\cdots,2)$.
The length of a vector $\a\in\N^d$ is defined as $|\a|\equiv\sum_{i=0}^{d-1}\a_{i}$.

Now given the homogenous structure of equation~\eqref{eq:160}, we can expand $f$ in polynomials 
and solve~\eqref{eq:160} among terms with a given multi-degree as multi-degree is invariant under permutations of 
$k_s$'s.  Thus without of loss of generality we can take $f$ to be homogeneous of multi-degree of $h$. 
Equation~(\ref{eq:160}) is then an equation for homogenous polynomials of multi-degree of $h + e_0$. 
Now the counterpart of~\eqref{eq:17} is 
\begin{equation}
f=k_{1}^{h}f_{0} +\sum_{a}^{\sim}k_{1}^{h-e_{\mu}}f_{e_{\mu}} +\sum_{\mu,\nu}^{\sim}k_{1}^{h-e_{\mu}-e_{\nu}}f_{e_{\mu}+e_{\nu}}+\cdots+f_{h}
\label{eq:35}
\end{equation}
where $f_{\a}$'s are homogeneous multi-symmetric polynomials of $k_2^\mu, \cdots k_n^\mu$ with multi-degree $\al$. 
Tilde over the sum means only summing over those cases where $h-\sum_{k}e_{\mu_{k}}\in\N^{d}$,
namely all components are nonnegative (we use the same notation below
unless specified).

With some manipulations we can rewrite~\eqref{eq:35} as (see Appendix~\ref{app:c4} for details)
\begin{equation} \label{qw}
f=\sum_{\eta}^{\sim}k_{1}^{\eta}g_{h-\eta}(q_{\alpha})+\sum_{\eta,\mu}^{\sim}k_{1}^{\eta-e_{\mu}}g_{h-\eta+e_{\mu}}(q_{\alpha})+\sum_{\eta,\mu,\nu}^{\sim}k_{1}^{\eta-e_{\mu}-e_{\nu}}g_{h-\eta+e_{\mu}+e_{\nu}}(q_{\alpha})+\cdots+g_{h}(q_{\alpha})
\end{equation}
which is the counterpart of~\eqref{eq:23-1}. In~\eqref{qw}  we should sum over all powers of $k_{1}$ with multi-degree $\eta$
with $|\eta|=n-1$ and $h-\eta\in\N^{d}$. $g_\ga$'s are homogeneous multi-symmetric polynomials of all $k^\mu_s$ with multi-degree $\ga$ and can be expanded in the multi-power sum basis
\begin{equation}
q_{\a}=\sum_{s=1}^{n}k_{s}^{\a}, \quad {\rm for \; all} \; \al \; {\rm with}  \quad 0\leq|\a|\leq n \ .
\end{equation}
Expanding $g_\ga$'s in terms of $\{q_\a\}$ and rearranging terms we can then write down the counterpart of~\eqref{eq:23} 
\begin{equation}
f=\sum_{P}^\sim \left(a^{(P)}_1  k_{1}^{\ga_{1}-e_0}q_{\ga_{2}}\cdots q_{\ga_{k}}+a^{(P)}_2 q_{\ga_{1}}k_{1}^{\ga_{2}-e_0}\cdots q_{\ga_{k}}+\cdots+a^{(P)}_k q_{\ga_{1}}\cdots q_{\ga_{k-1}}k_{1}^{\ga_{k}-e_0}\right)
\label{eq:42}
\end{equation}
where $P$ denotes a partition of $h + e_0$ (i.e. $\sum_{i=1}^k \ga_i = h+e_0$) and we sum over all partitions.
Note that by definition $\ga_i - e_0 \in \N^d$ for all $i$. 
We can again separate the summation over $P$ into those which have at least one $|\ga_j| =1$ and those with none (which we denote
as $P'$). The former is either zero by momentum conservation or fully symmetric in all the $k_s^\mu$ (if some $\ga_j = e_0$), and thus 
\be 
f = f^{(s)} + f^{(a)}, \qquad f^{(a)} =  \sum_{P'} \le(\cdots \ri)
\ee
where $f^{(s)}$ is fully symmetric in all $k_s^\mu$ and in $P'$ all $|\ga_j| \geq 2$. 

Now~\eqref{eq:160} gives  
\be 
\sum_{P'}  \le(a^{(P')}_1 + a^{(P')}_2 + \cdots + a^{(P')}_k \ri) 
q_{\ga_{1}}q_{\ga_{2}}\cdots q_{\ga_{k}}  = 0 
\ee
and then
\be  \label{sol1}
- \le(a^{(P')}_1 + a^{(P')}_2 + \cdots + a^{(P')}_{k-1} \ri) = a^{(P')}_k 
\ee
for all partitions $P'$. Substituting the above equation back into $f^{(a)}$ we again have 
\begin{equation}
\sI^{(a)}_b = \sum_{P'} \sum_{j=1}^{k-1}a^{(P')}_j q_{\ga_{1}}\cdots\cancel{q_{\ga_{j}}}\cdots q_{\ga_{k-1}}(k_{1}^{\ga_{j}-e_0}q_{\ga_{k}}-q_{\ga_{j}}k_{1}^{\ga_{k}-e_0}) \chi_a \chi_r^{n-1}
\end{equation}
and using permutations among $\chi_r$'s
we then replace  $k_{1}^{\ga_{j}-e_0}q_{\ga_{k}}-q_{\ga_{j}}k_{1}^{\ga_{k}-e_0}$ by
\begin{align} \label{ii}
 (n-1) \le( k_{1}^{\g_{j}-e_{0}} k_2^{\g_{k}}-k_2^{\g_{j}}k_{1}^{\g_{k}-e_{0}}\ri)
= \om_2  (n-1) \le( k_{1}^{\g_{j}-e_{0}} k_2^{\g_{k}-e_0}-k_2^{\g_{j}-e_0}k_{1}^{\g_{k}-e_{0}}\ri) \ .
\end{align}
Thus $\sI^{(a)}_b$ again has the structure of~\eqref{uei}.

\subsection{General case}

We now consider the general case where $\chi$ can have arbitrary species indices. 
The idea is exactly the same as before except that now we view~\eqref{ibb} as a multi-variable polynomial 
of both momenta and fields. More explicitly, define 
\bega 
\bar k_s \equiv (k_s^\mu, \xi_s^\rmi), \quad \xi_s^\rmi \equiv \chi_{r \rmi} (k_s) \quad s=1, \cdots n , \quad \rmi =1, \cdots, N \\
 \bar k_{1a} = (k_1^\mu, {\xi}_a^\rmi ), \qquad \xi_a^\rmi \equiv \chi_{a \rmi} (k_1)
\end{gather}
as $d+N$ dimensional vectors. Then $\sI_b$ of~\eqref{ibb} (when we expand all the coefficients in momenta) can be viewed as a polynomial $\sI_b (\bar k_{1a}; \bar k_2, \cdots , \bar k_n)$ which is symmetric in the last $n-1$ variables,  and equation~\eqref{skm3} (after symmetrization) becomes 
\begin{equation}
\w_{1}\sI_{b}(\bar{k}_{1};\bar{k}_{2},\cdots,\bar{k}_{n})+\w_{2} \sI_{b}(\bar{k}_{2};\bar{k}_{1},\cdots,\bar{k}_{n})+\cdots+\w_{n} \sI_{b}(\bar{k}_{n};\bar{k}_{1},\cdots,\bar{k}_{n-1})=0 \ .
\label{eq:mostkms}
\end{equation}
The story is similar to before and we can solve~\eqref{eq:mostkms} for $\sI_b$ with a given multi-degree $h$ which is now 
a $d+N$-dimensional vector. There are two important differences: 

\ben 

\item Momentum conservation only applies to the first $d$ components of $\bar k$. We thus separate the multi-agree for $\bar k$ as $\ga = (\hat \ga, \tilde \ga)$ where $\hat \ga$ is $d$-dimensional and $\tilde \ga$ is $N$-dimensional, i.e. 
$\bar k^\ga = k^{\hat \ga} \xi^{\tilde \ga}$. Similarly the total multi-degree $h = (\hat h, \tilde h)$ and the basis are $\hat e_a$ with $a =0,1, \cdots d-1$ and $\tilde e_\rmi$ with $\rmi = 1, \cdots N$. 

\item We only need to consider polynomials of the form~\eqref{ibb}, i.e. $|\tilde \ga_s|$ for each $\bar k_s$ can only be $1$ or $0$ and 
$|\tilde h|  = n$. 
\een 

We can now expand $\sI_b (\bar k_1; \bar k_2 , \cdots \bar k_n)$ in the form of~\eqref{eq:42} with $q_\al$ now defined in terms of $\bar k$, $k_1$ replaced by $\bar k_1$, and $e_0$  replaced by $\hat e_0$. In particular for any $i$ we should have 
$\hat \ga_i - \hat e_0 \in \N^d$. We still have 
\be 
\sI_b = \sI_b^{(s)} + \sI_b^{(a)}, \qquad \sI_b^{(a)} = \sum_{P'} (\cdots) 
\ee
and~\eqref{sol1} with the only difference that $P'$ also includes those partitions with $|\ga_i| \geq 2$ for all $i$ {\it and} those with $|\ga_i| = |\tilde \ga_i |=1$.  
Now plugging~\eqref{sol1} into $\sI_b^{(a)}$ and replacing $\bar k_1$ by $\bar k_{1a}$ we  find that 
\begin{equation} \label{90}
\sI^{(a)}_b = \sum_{P'} \sum_{i=1}^{k-1}a^{(P')}_i q_{\ga_{1}}\cdots\cancel{q_{\ga_{i}}}\cdots q_{\ga_{k-1}}(\bar k_{1a}^{\ga_{i}-\hat e_0}q_{\ga_{k}}-q_{\ga_{i}} \bar k_{1a}^{\ga_{k}-\hat e_0})
\end{equation}
where we should also replace all the $\bar k_1$'s in $q_\ga$'s by $\bar k_{1a}$. From item 2 discussed earlier
the terms relevant for $\sI_b$  must have $|\tilde{\g}_{i}|=1$ or $0$ for all
$i$'s and $\sum_{i=1}^k |\tilde \ga_i| =n$.  For a $\ga_i$ with $|\tilde{\g}_{i}|=1$ we will then have  
$\tilde \ga_i = \tilde e_{\rmj_i}$ for some $\rmj_i$.  Those with $|\tilde{\g}_{i}|=0$ can be written as $q_{\g_{i}}= {q}_{\hat{\g}_{i}}\equiv\sum_{s=1}^{n}k_{s}^{\hat{\g}_{i}}$.  

Since for all $\ga_i - \hat e_0 \in \N^d$, by exchanging indices in~\eqref{90} we can always extract an overall factor $\om$ and the remaining factor in~\eqref{90} is then anti-symmetric between $\bar k_{1a}$ and $\bar k_2$, thus giving~\eqref{uei}.
To see this explicitly  we must keep in mind that we should only select those terms in~\eqref{90} which has the structure of~\eqref{ibb}.\footnote{All the rest terms must cancel themselves by definition.}
There are three types of terms in~\eqref{90}:

\ben 

\item $|\tilde \ga_i| = |\tilde \ga_k |=1$ for which we have 
\bea 
\bar k_{1a}^{\ga_{i}-\hat e_0}q_{\ga_{k}}-q_{\ga_{i}} \bar k_{1a}^{\ga_{k}-\hat e_0}
&= & k_1^{\hat \ga_i - \hat e_0} \xi_{1a}^{\rmj_i} \sum_{s=2}^n k_s^{\hat \ga_k} \xi_{s}^{\rmj_k}  
-  k_1^{\hat \ga_k - \hat e_0} \xi_{1a}^{\rmj_k}\sum_{s=2}^n k_s^{\hat \ga_i} \xi_{s}^{\rmj_i}   \cr
&=&  (n-1)\om \le(  k_1^{\hat \ga_i - \hat e_0} \xi_{1a}^{\rmj_i}  k_2^{\hat \ga_k - \hat e_0} \xi_{2}^{\rmj_k}  
-  k_1^{\hat \ga_k - \hat e_0} \xi_{1a}^{\rmj_k}  k_2^{\hat \ga_i-\hat e_0} \xi_{2}^{\rmj_i}  \ri)
\label{oop}
\eea
where in the second line we have used that in~\eqref{90} the expression is multiplied by expressions symmetric in $\bar k_2 , \cdots \bar k_n$. Equation~\eqref{oop} leads to terms in~\eqref{90}  of the form~\eqref{uei}. 

\item   $|\tilde \ga_i| =1$ and $|\tilde \ga_k |=0$ for which we have 
\bea 
\bar k_{1a}^{\ga_{i}-\hat e_0}q_{\ga_{k}}-q_{\ga_{i}} \bar k_{1a}^{\ga_{k}-\hat e_0}
&= & k_1^{\hat \ga_i - \hat e_0} \xi_{1a}^{\rmj_i} \sum_{s=2}^n k_s^{\hat \ga_k} 
-  k_1^{\hat \ga_k - \hat e_0} \sum_{s=2}^n k_s^{\hat \ga_i} \xi_{s}^{\rmj_i}   
\cr
&=&  (n-1)\om \le(  k_1^{\hat \ga_i - \hat e_0} \xi_{1a}^{\rmj_i}  k_2^{\hat \ga_k - \hat e_0} 
-  k_1^{\hat \ga_k - \hat e_0}  k_2^{\hat \ga_i-\hat e_0} \xi_{2}^{\rmj_i}  \ri)\ .
\eea
Now the two terms on the right hand side must select different factors $\xi_s$ from the product of $q_j$'s in~\eqref{90} so that they will have the structure of~\eqref{ibb}. %It is clear that given the anti-symmetric structure of~\eqref{oop}, the resulting term 
Take some  $|\tilde \ga_l| = 1$, then $q_{\ga_l} \le(\bar k_{1a}^{\ga_{i}-\hat e_0}q_{\ga_{k}}-q_{\ga_{i}} \bar k_{1a}^{\ga_{k}-\hat e_0}\ri)$ will result in terms
\be
%k_1^{\hat \ga_i - \hat e_0} \xi_{1a}^{\rmj_i} k_2^{\hat \ga_k} k_2^{\hat \ga_l} \xi_2^{\rmj_l} 
%-  k_1^{\hat \ga_k - \hat e_0}  k_1^{\hat \ga_l} \xi_{1a}^{\rmj_l}   k_2^{\hat \ga_i} \xi_{2}^{\rmj_i}  
(n-1) \om \le(k_1^{\hat \ga_i - \hat e_0}  k_2^{\hat \ga_k + \hat \ga_l - \hat e_0}  \xi_{1a}^{\rmj_i} \xi_2^{\rmj_l} 
-   k_1^{\hat \ga_k + \hat \ga_l - \hat e_0} k_2^{\hat \ga_i-\hat e_0} \xi_{1a}^{\rmj_i}   \xi_{2}^{\rmj_l}    \ri)
\ee
which again have the structure of~\eqref{uei}. 

\item $|\tilde \ga_i| = |\tilde \ga_k |=0$: this is case is similar to~\eqref{ii} and we have 
\be
\bar k_{1a}^{\ga_{i}-\hat e_0}q_{\ga_{k}}-q_{\ga_{i}} \bar k_{1a}^{\ga_{k}-\hat e_0} =
  \om_2  (n-1) \le( k_{1}^{\hat \g_{j}-e_{0}} k_2^{\hat \g_{k}-e_0}-k_2^{\hat \g_{j}-e_0}k_{1}^{\hat \g_{k}-e_{0}}\ri) \ .
 \ee
Multiplying it with products of $q_i$'s in~\eqref{90} and selecting the appropriate terms for~\eqref{ibb} we always 
find terms of the form~\eqref{uei}, due to full permutation symmetry of $q_i$'s.

\een 
This then concludes the full proof.

\subsection{Justification of expansions using symmetric polynomials} \label{app:c4}

We now show that~\eqref{eq:23-1} follows from~\eqref{eq:17}. For this purpose we can expand $f_m$ in~\eqref{eq:17}
in power sum basis (the fundamental theorem
of symmetric polynomials) of $\om_2, \cdots \om_n$
\begin{equation}
p_{k}=\sum_{s=2}^{n}\w_{s}^{k},\qquad(k=0,\cdots,n-1)
\end{equation}
From~\eqref{qk} we have $q_{k}=\w_{1}^{k}+p_{k}$.
Plugin this relation into (\ref{eq:17}) and we can rewrite $f$ in
terms of the following expansion:
\begin{equation}
f=\w_{1}^{h}\tilde{g}_{0}(q_{i})+\w_{1}^{h-1}\tilde{g}_{1}(q_{i})+\cdots+\w_{1}\tilde{g}_{h-1}(q_{i})+\tilde{g}_{h}(q_{i})\label{eq:19}
\end{equation}
where $\tilde{g}_{i}$'s are polynomials of degree $i$ and of $q_{k}$
. Note this expansion in terms of $q_{k}$ and $\w_{1}$ is not unique
although the expansion (\ref{eq:17}) is. Define the symmetric polynomial
basis of $\w_{1}$ to $\w_{n}$ as
\begin{equation}
\s_{i}=\sum_{1\leq k_{1}<\cdots<k_{i}\leq n}\w_{k_{1}}\cdots\w_{k_{i}},\;\s_{0}=1
\end{equation}
for $(i=0,\cdots,n)$ and that of $\w_{2}$ to $\w_{n}$ as $s_{i}$
for $(i=0,\cdots,n-1)$. Using the following relations
\begin{align}
\s_{i} & =\w_{1}s_{i-1}+s_{i},\qquad i=1,\cdots,n-1\\
\s_{n} & =\w_{1}s_{n-1}
\end{align}
we can expand $\w_{1}^{n+a}$ in the form of (\ref{eq:19}) in terms
lower powers of $\w_{1}$ when $a\geq0$:
\begin{equation}
\w_{1}^{n+a}=\w_{1}^{a}(\w_{1}^{n-1}\s_{1}-\w_{1}^{n-2}\s_{2}+\cdots-(-1)^{n}\s_{n})\label{eq:23-2}
\end{equation}
This expansion is equivalent to (\ref{eq:19}) because $\s_{i}$ and
$q_{i}$ can be uniquely expanded by each other. Following this process,
one can always reduce $h$ in (\ref{eq:19}) to be less than $n$
and~\eqref{eq:23-1} results.

We now justify~\eqref{qw}. 
Like symmetric polynomials, any multisymmetric polynomial can be uniquely
expanded by elementary multisymmetric polynomials
\begin{equation}
\s_{\alpha}=\sum_{\substack{\{i_{j,k}\}\\
1\leq i_{j,k}\leq n,\text{ all different}
}
}\w_{i_{0,1}}\cdots\w_{i_{0,\alpha_{0}}}k_{i_{1,1}}^{1}\cdots k_{i_{1,\a_{1}}}^{1}\cdots k_{i_{d-1,1}}^{d-1}\cdots k_{i_{d-1,\a_{d-1}}}^{d-1}
\end{equation}
for all posible choices of $\a$ with $0\leq|\a|\leq n$. We define
$\s_{\a}$ for $k_{1}$ to $k_{n}$ and $s_{\a}$ for $k_{2}$ to
$k_{n}$. For example,
\begin{equation}
\s_{(2,1,1)}=\sum_{\substack{i_{1}<i_{2}\\
1\leq i_{j}\leq n,\text{ all different}
}
}\w_{i_{1}}\w_{i_{2}}k_{i_{3}}^{1}k_{i_{4}}^{2}
\end{equation}

We define
$q_{\a}$ for $k_{1}$ to $k_{n}$ and $p_{\a}$ for $k_{2}$ to $k_{n}$.
For example,
\begin{equation}
q_{(2,1,1)}=\sum_{i=1}^{n}\w_{i}^{2}k_{i}^{1}k_{i}^{2}
\end{equation}
These two basis can be expanded via each other uniquely. As a necessary
condition, by the notation of $d$-vector $\a$, we see the number
of these two basis are the same.

Expanding $f_\ga$ in~\eqref{eq:35} in terms of $p_{\alpha}$ and using the relations
\begin{equation}
q_{\a}=k_{1}^{\a}+p_{\a}
\end{equation}
we can rewrite (\ref{eq:35}) in the form of
\begin{equation}
f=k_{1}^{h}\tilde{g}_{0}(q_{\alpha})+\sum_{j}^{\sim}k_{1}^{h-e_{j}}\tilde{g}_{e_{j}}(q_{\a})+\sum_{j,l}^{\sim}k_{1}^{h-e_{j}-e_{l}}\tilde{g}_{e_{j}+e_{l}}(q_{\alpha})+\cdots+\tilde{g}_{h}(q_{\alpha})\label{qe:37}
\end{equation}
Furthermore, by the relations
\begin{align}
\s_{\a} & =\sum_{i}^{\sim}k_{1}^{i}s_{\a-e_{i}}+s_{\a},\;0\leq|\a|\leq n-1\\
\s_{\a} & =\sum_{i}^{\sim}k_{1}^{i}s_{\a-e_{i}},\;|\a|=n
\end{align}
we have a generalized version of identity (\ref{eq:23-2}):
\begin{equation}
k_{1}^{\a+\b}=\f{k_{1}^{\b}}{n!}\sum_{j=0}^{n-1}\sum_{\{i_j\}}^{\sim}(-1)^{n-j+1}k_{1}^{e_{i_1}+\cdots+e_{i_j}}\s_{\a-(e_{i_1}+\cdots+e_{i_j})}
\end{equation}
for any $|\a|=n$ and $|\b|\geq0$. Using this method, we can reduce
(\ref{qe:37}) to an expansion with the length of maximal exponent
of $k_{1}$ to be less than $n$ and gives~\eqref{qw}.

\end{document}

%% file: HLdefs.tex
\usepackage{color}

\newcommand{\nb}[1]{\color{blue}}

\newcommand{\hl}[1]{\color{magenta}}

\usepackage[
	  pagebackref=false,
	  colorlinks=true,
      linkcolor=blue,
      urlcolor=blue,
      filecolor=black,
      citecolor=red,
      pdfstartview=FitV,
      pdftitle={},
        pdfauthor={Thomas Faulkner, Hong Liu, Mukund Rangamani},
        pdfsubject={},
        pdfkeywords={},
        pdfpagemode=None,
        bookmarksopen=true
      ]{hyperref}

\usepackage[normalem]{ulem}
\usepackage{amsmath}
\usepackage{enumerate}
\usepackage{amsfonts}
\usepackage{epsfig}
\usepackage{mathbbol}
\usepackage{cancel}
\usepackage{mathrsfs}

\def\Tr{\mathop{\rm Tr}}

\newcommand\half{{\ensuremath{\frac{1}{2}}}}
\newcommand\p{\ensuremath{\partial}}

\newcommand{\be}{\begin{equation}}
\newcommand{\ee}{\end{equation}}
\newcommand{\bea}{\begin{eqnarray}}
\newcommand{\eea}{\end{eqnarray}}
\newcommand{\bega}{\begin{gather}}
\newcommand{\eega}{\end{gather}}

\newcommand{\bi}{\begin{itemize}}
\newcommand{\ei}{\end{itemize}}
\newcommand{\ben}{\begin{enumerate}}
\newcommand{\een}{\end{enumerate}}
\newcommand{\bca}{\begin{cases}}
\newcommand{\eca}{\end{cases}}
\newcommand{\bln}{\begin{align}}
\newcommand{\eln}{\end{align}}
\newcommand{\bst}{\begin{split}}
\newcommand{\est}{\end{split}}
\def\ie{\begin{equation}\begin{aligned}}
\def\fe{\end{aligned}\end{equation}}
\newcommand{\bma}{\le(\begin{matrix}}
\newcommand{\ema}{\end{matrix}\ri)}

\newcommand\al{{\alpha}}
\newcommand\ep{\epsilon}
\newcommand\sig{\sigma}
\newcommand\Sig{\Sigma}
\newcommand\lam{\lambda}
\newcommand\Lam{\Lambda}
\newcommand\om{\omega}

\newcommand\ga{{\ensuremath{{\gamma}}}}

\newcommand\de{{\ensuremath{{\delta}}}}

\newcommand\vp{\varphi}

\newcommand\da{{\dagger}}

\def\th{{\theta}}

\newcommand\ra{{\rightarrow}}

\newcommand\ov{\over}
\newcommand\ha{{\half}}

\def\le{\left}
\def\ri{\right}

\newcommand\sA{{\ensuremath{{\mathcal A}}}}
\newcommand\sB{{\ensuremath{{\mathcal B}}}}
\newcommand\sC{{\ensuremath{{\mathcal C}}}}

\newcommand\sE{{\ensuremath{{\mathcal E}}}}
\newcommand\sF{{\ensuremath{{\mathcal F}}}}
\newcommand\sI{{\ensuremath{{\mathcal I}}}}
\newcommand\sG{{\ensuremath{{\mathcal G}}}}
\newcommand\sH{{\ensuremath{{\mathcal H}}}}

\newcommand\sL{{\ensuremath{{\mathcal L}}}}

\newcommand\sP{{\ensuremath{{\mathcal P}}}}

\newcommand\sT{{\mathcal T}}

\newcommand\vx{{\vec x}}
\newcommand\vk{{\vec k}}

\newcommand{\hmu}{{\hat \mu}}

\newcommand{\rmi}{{\rm i}}
\newcommand{\rmj}{{\rm j}}
\newcommand{\rmk}{{\rm k}}

\global\long\def\mA{\mathcal{A}}
\global\long\def\mB{\mathcal{B}}

\global\long\def\mF{\mathcal{F}}

\global\long\def\mH{\mathcal{H}}

\global\long\def\mL{\mathcal{L}}
\global\long\def\mM{\mathcal{M}}

\global\long\def\mX{\mathcal{X}}

\global\long\def\e{\epsilon}
\global\long\def\ra{\rightarrow}

\global\long\def\f#1#2{\frac{#1}{#2}}
\global\long\def\del{\partial}
\global\long\def\t{\theta}
\global\long\def\a{\alpha}
\global\long\def\b{\beta}
\global\long\def\g{\gamma}
\global\long\def\G{\Gamma}
\global\long\def\s{\sigma}

\global\long\def\d{\delta}
\global\long\def\Tr{\text{Tr}}

\global\long\def\N{\mathbb{N}}

\global\long\def\w{\omega}

%% file: susy_draft_8.bbl
\begin{thebibliography}{9}


\bibitem{schwinger} J. Schwinger, J. Math. Phys. {\bf 2}, 407 (1961); Particles and Sources, vol. I., II., and III., Addison-Wesley, Cambridge, Mass. 1970-73.

\bibitem{keldysh}  L. V. Keldysh, Zh. Eksp. Teor. Fiz. {\bf 47}, 1515 (1964) (Sov. Phys. JETP {\bf 20}, 1018 (1965)).

%\cite{Feynman:1963fq}
\bibitem{Feynman:1963fq} 
  R.~P.~Feynman and F.~L.~Vernon, Jr.,
  %``The Theory of a general quantum system interacting with a linear dissipative system,''
  Annals Phys.\  {\bf 24}, 118 (1963)
  [Annals Phys.\  {\bf 281}, 547 (2000)].
  %%CITATION = doi:10.1016/0003-4916(63)90068-X;%%


%\cite{Chou:1984es}
\bibitem{Chou:1984es}
  K.~C.~Chou, Z.~B.~Su, B.~L.~Hao and L.~Yu,
  %``Equilibrium and Nonequilibrium Formalisms Made Unified,''
  Phys.\ Rept.\  {\bf 118}, 1 (1985).
  %%CITATION = PRPLC,118,1;%%

\bibitem{Wang:1998wg}
  E.~Wang and U.~W.~Heinz,
  ``A Generalized fluctuation dissipation theorem for nonlinear response functions,''
  Phys.\ Rev.\ D {\bf 66}, 025008 (2002)
  [hep-th/9809016].
  %%CITATION = HEP-TH/9809016;%%


%\cite{Glorioso:2016gsa}
\bibitem{GL} 
  P.~Glorioso and H.~Liu,
  %``The second law of thermodynamics from symmetry and unitarity,''
  arXiv:1612.07705 [hep-th].
  %%CITATION = ARXIV:1612.07705;%%
  
  %\cite{Crossley:2015evo}
\bibitem{CGL}
  M.~Crossley, P.~Glorioso and H.~Liu,
  ``Effective field theory of dissipative fluids,''
  arXiv:1511.03646 [hep-th].
  %%CITATION = ARXIV:1511.03646;%%

\bibitem{CGL1}
 P.~Glorioso, M.~Crossley, and H.~Liu, %``Effective field theory of dissipative fluids II,'' to appear.
%\cite{Glorioso:2017fpd}
%\bibitem{Glorioso:2017fpd} 
%  P.~Glorioso, M.~Crossley and H.~Liu,
  ``Effective field theory for dissipative fluids (II): classical limit, dynamical KMS symmetry and entropy current,''
  arXiv:1701.07817 [hep-th].
  %%CITATION = ARXIV:1701.07817;%%

\bibitem{Grozdanov:2013dba}
S.~Grozdanov and J.~Polonyi, %``{Viscosity and dissipative hydrodynamics from effective field theory},''
arXiv:1305.3670 [hep-th].
%%CITATION = ARXIV:1305.3670;%%.


  %\cite{Kovtun:2014hpa}
\bibitem{Kovtun:2014hpa}
  P.~Kovtun, G.~D.~Moore and P.~Romatschke,
  %``Towards an effective action for relativistic dissipative hydrodynamics,''
  JHEP {\bf 1407} (2014) 123
  [arXiv:1405.3967 [hep-ph]].
  %%CITATION = ARXIV:1405.3967;%%


%\cite{Kovtun:2014hpa,Harder:2015nxa}
\bibitem{Harder:2015nxa}
  M.~Harder, P.~Kovtun and A.~Ritz,
  %``On thermal fluctuations and the generating functional in relativistic hydrodynamics,''
  JHEP {\bf 1507}, 025 (2015)
  [arXiv:1502.03076 [hep-th]].
  %%CITATION = ARXIV:1502.03076;%%




\bibitem{Haehl:2014zda}
   F.~M.~Haehl, R.~Loganayagam and M.~Rangamani,
  %``Adiabatic hydrodynamics: The eightfold way to dissipation,''
  JHEP {\bf 1505}, 060 (2015)
  [arXiv:1502.00636 [hep-th]].
  %%CITATION = ARXIV:1502.00636;%%


\bibitem{Haehl:2015foa}
  F.~M.~Haehl, R.~Loganayagam and M.~Rangamani,
  %``The Fluid Manifesto: Emergent symmetries, hydrodynamics, and black holes,''
  arXiv:1510.02494 [hep-th].
  %%CITATION = ARXIV:1510.02494;%%


%\cite{Haehl:2015foa,Haehl:2015uoc}
\bibitem{Haehl:2015uoc} 
  F.~M.~Haehl, R.~Loganayagam and M.~Rangamani,
  %``Topological sigma models & dissipative hydrodynamics,''
  JHEP {\bf 1604}, 039 (2016)
  [arXiv:1511.07809 [hep-th]]; 
  
  

\bibitem{Sieberer1} L. M. Sieberer, M. Buchhold, and S. Diehl, %``Keldysh Field Theory for Driven Open Quantum Systems,''
Rep. Prog. Phys. {\bf 79}, 096001 (2016); arXiv:1512.00637.




\bibitem{msr} P. C. Martin, E. D. Siggia and H. A. Rose. Phys. Rev. {\bf A8}, 423 (1973).

\bibitem{Dedo} J. DeDominicis. J. Physique (Paris) {\bf 37}, C1 (1976).


\bibitem{janssen1} H. Janssen, Z. Phys.  {\bf B23}, 377 (1976).

\bibitem{Kamenev} A. Kamenev, ``Field Theory of Non-Equilibrium Systems,'' Cambridge University Press, Cambridge (2011).






\bibitem{zinnjustin} J.~Zinn-Justin, ``Quantum Field Theory and Critical Phenomena,'' Clarendon Press, Oxford (2002).





%\cite{Haehl:2016pec}
\bibitem{Haehl:2016pec}
  F.~M.~Haehl, R.~Loganayagam and M.~Rangamani,
  %``Schwinger-Keldysh formalism I: BRST symmetries and superspace,''
  arXiv:1610.01940 [hep-th].
  %%CITATION = ARXIV:1610.01940;%%
 

\bibitem{kubo57} R. Kubo, %``Statistical mechanical theory of irreversible processes I,''
J. Math. Soc. Japan {\bf 12} 570 (1957).

\bibitem{mart59} P. C. Martin and J. Schwinger, %``Theory of many particle systems I,''
Phys. Rev. {\bf 115} 1342 (1959).

\bibitem{Kadanoff}  L. P. Kadanoff and P. C. Martin, %``Hydrodynamic equations and correlation functions,''
Ann. Phys. {\bf 24}, 419 (1963).


\bibitem{janssen2} R. Bausch, H. K. Janssen, and H. Wagner, Z. Phys. {\bf B24}, 113 (1976).

\bibitem{Sieberer2} L. M. Sieberer, A. Chiocchetta, A. Gambassi, U. C. Tauber, and S.~Diehl, %``Thermodynamic Equilibrium as a Symmetry of the Schwinger-Keldysh Action,''
Phys. Rev. {\bf B 92} 134307  (2015);  arXiv:1505.00912.



%\cite{Niemi:1983nf}
%\bibitem{Niemi:1983nf}
%  A.~J.~Niemi and G.~W.~Semenoff,
  %``Finite Temperature Quantum Field Theory in Minkowski Space,''
 % Annals Phys.\  {\bf 152}, 105 (1984).
  %%CITATION = APNYA,152,105;%%

%\cite{Chou:1984es,Wang:1998wg}

\bibitem{parisi} G. Parisi and N. Sourlas, %``Random Magnetic Fields, Supersymmetry and Negative
%Dimensions,''
Phys. Rev. Lett. {\bf 43} 744 (1979).

\bibitem{feigelman}  M. Feigelman and A. Tsvelik, %``On the Hidden Supersymmetry of Fokker-Planck
%Equations with Potential Forces,''
Phys. Lett. {\bf A95} 469 (1983).


%\cite{parisi,feigelman,Gozzi:1983rk,Mallick:2010su,zinnjustin,Haehl:2015foa}
\bibitem{Gozzi:1983rk}
  E.~Gozzi,
  %``The Onsager's Principle of Microscopic Reversibility and Supersymmetry,''
  Phys.\ Rev.\ D {\bf 30}, 1218 (1984)
 % [Phys.\ Rev.\ D {\bf 31}, 441 (1985)].
  %%CITATION = PHRVA,D30,1218;%%

%\cite{Mallick:2010su}
\bibitem{Mallick:2010su}
  K.~Mallick, M.~Moshe and H.~Orland,
  %``A Field-theoretic approach to nonequilibrium work identities,''
  J.\ Phys.\ A {\bf 44}, 095002 (2011)
  [arXiv:1009.4800 [cond-mat.stat-mech]].
  %%CITATION = ARXIV:1009.4800;%%

%Haehl:2016pec
%\cite{Haehl:2016pec,Haehl:2016uah}
\bibitem{Haehl:2016uah} 
  F.~M.~Haehl, R.~Loganayagam and M.~Rangamani,
  %``Schwinger-Keldysh formalism II: Thermal equivariant cohomology,''
  arXiv:1610.01941 [hep-th].
  %%CITATION = ARXIV:1610.01941;%%

\bibitem{yarom}  K.~Jensen, N.~Pinzani-Fokeeva, and A.~Yarom, 
``Dissipative hydrodynamics in superspace,'' 
  arXiv:1701.07436 [hep-th].
  %%CITATION = ARXIV:1701.07436;%%
  %1 citations counted in INSPIRE as of 20 Feb 2017


\bibitem{hohenberg} P. C. Hohenberg and B. I. Halperin, Rev. Mod. Phys. {\bf 49}, 435
(1977).

\bibitem{Folk} R. Folk and G. Moser, J. Phys. {\bf A39}, R207 (2006).




\bibitem{Bagger1992}
J.~Bagger,
\newblock {\em Supersymmetry and supergravity}.
\newblock Princeton University Press, 1992.

\bibitem{briand2004algebra}
E.~Briand,
%\newblock When is the algebra of multisymmetric polynomials generated by the
%  elementary multisymmetric polynomials?
\newblock {\rm Contributions to Algebra and Geometry} {\bf 45}, 353 (2004).

\bibitem{dalbec1999multisymmetric}
J.~Dalbec,
%\newblock Multisymmetric functions.
\newblock {\rm Beitr{\"a}ge Algebra Geom} {\bf 40}, 27 (1999).










\end{thebibliography}
